\documentclass[]{emulateapj} 
\usepackage[colorlinks=true,urlcolor=blue,citecolor=blue,linkcolor=blue]{hyperref}
\usepackage{graphicx}
\usepackage{url}
\usepackage[usenames]{color}
\usepackage{natbib}
\usepackage{amsmath}

\def\edt#1{#1}

\def\para{\mathrm{para}}
\def\CaII{\ion{Ca}{2}}
\def\MgII{\ion{Mg}{2}}
\def\HeI{\ion{He}{I}}
\def\HeI1083{\HeI\ 1083 nm}
\def\Bifrost{{\it Bifrost}}

\def\multitd{{\it Multi3d}}

\def\figspath{.}
\def\Cair{\CaII\ 854.2 nm}
\def\Halpha{\mbox{H$\alpha$}}
\def\kms{\mbox{km s$^{-1}$}}
\newcommand{\dd}{\mathrm{d}}
\newcommand{\be}{\begin{equation}}
\newcommand{\ee}{\end{equation}}
\newcommand{\vect}[1]{\boldsymbol{#1}}
\newcommand{\bea}{\begin{eqnarray}}
\newcommand{\eea}{\end{eqnarray}}
\begin{document}

\title{On fibrils and field lines: The nature of H$\alpha$ fibrils in the solar
chromosphere}
  
   \author{Jorrit Leenaarts$^{1}$}\email{jorrit.leenaarts@astro.su.se}
   \author{Mats Carlsson$^{2}$}\email{mats.carlsson@astro.uio.no}
   \author{Luc Rouppe van der Voort$^{2}$}\email{rouppe@astro.uio.no}
 
\affil{$^1$ Institute for Solar Physics, Department of Astronomy,
  Stockholm University,
AlbaNova University Centre, SE-106 91 Stockholm Sweden}
\affil{$^2$ Institute of
  Theoretical Astrophysics, University of Oslo, P.O. Box 1029
  Blindern, N--0315 Oslo, Norway}

\date{Received; accepted}

\begin{abstract}
  Observations of the solar chromosphere in the line-core of the
  \Halpha\ line show dark elongated structures called fibrils that
  show swaying motion. We performed a 3D radiation-MHD simulation of a
  network region, and computed synthetic \Halpha\ images from this
  simulation to investigate the relation between fibrils and the
  magnetic field lines in the chromosphere. The synthetic imagery
  shows fibrils just as in observations. The periods, amplitudes and
  phase-speeds of the synthetic fibrils are consistent with those
  observed. We analyse the relation between the synthetic fibrils and
  the field lines threading through them, and find that some fibrils
  trace out the same field line along the fibril's length, but there
  are also fibrils that sample different field lines at different
  locations along their length. Fibrils sample the same field lines on
  a time scale of $\sim200$~s. This is shorter than their own
  lifetime, and thus they do not sample the same field lines during
  their entire life. We analysed the evolution of the atmosphere along
  a number of field lines that thread through fibrils and find that
  they carry slow-mode waves that load mass into the field line, as
  well as transverse waves that propagate with the Alfv\'en
  speed. Transverse waves propagating in opposite directions cause an
  interference pattern with complex apparent phase speeds.  These wave
  motions are present on top of a slow drift of the field line through
  the computational domain.
 
The relationship between fibrils and field lines is complex. It is
governed by constant migration and swaying of the field lines, their
mass loading by slow modes and subsequent draining, and their actual
visibility in \Halpha.  Field lines are visible where they lie close
to the optical depth unity surface. The location of the latter is
governed by the height at which the column mass in the chromosphere
reaches a certain value. The visibility of field line at a fixed time
is thus not only governed by its own mass density but also by the mass
density of the material above it. We conclude that using the swaying
motion of fibrils as a tracer of chromospheric transverse oscillations
must be done with caution.
     
\end{abstract}

   \keywords{Sun: atmosphere --- Sun: chromosphere --- Sun: oscillations ---
     magnetohydrodynamics (MHD)}
  
%_______________________________________________________________________
\section{Introduction}                          \label{sec:introduction}
%_______________________________________________________________________

Images of the solar disk taken in the line core of the \Halpha\ line display
dramatic elongated dark and bright features called fibrils and
mottles. They are present everywhere, except in the quietest Sun away
from concentrations of magnetic field. Their elongated shape and
presence wherever the magnetic field is strong, suggest that
fibrils and mottles outline the plane-of-the-sky component of the
magnetic field.

There is, however, very little direct observational evidence for this
suggestion in network areas and quiet Sun regions.  The \Halpha\ line
itself is ill-suited for measuring chromospheric magnetic fields
\citep{2004ApJ...603L.129S}.
The plane-of-the sky magnetic field around sunspots was measured by
\citet{2011A&A...527L...8D}
using spectropolarimetry in the \Cair\ line. This line has a similar
formation height as \Halpha\
\citep{2013ApJ...772...90L}
and shows similar fibril structure. Their conclusion is that most of
the time, but not always, fibrils appear to be aligned with the
plane-of-the sky magnetic field. To our knowledge no studies in quiet
Sun have been done so far.

On the modelling side only
\citet{2012ApJ...749..136L}
have reproduced fibril-like structures in synthetic \Halpha\ images
computed from a snapshot of a radiation-MHD (RMHD) simulation of the
solar atmosphere. They found that the fibril-like structures in the
simulation indeed trace out the horizontal component of the magnetic
field, but did not measure the alignment of the 3D magnetic field
vector and the fibril-like structure. It is thus not clear whether
fibrils trace out single field lines, or that they actually intersect
a set of different field lines that happen to have a horizontal
component parallel to the fibril.

For a subset of dark elongated \Halpha\ features, there is compelling
evidence based on their dynamical behaviour that they align with the
magnetic field in all three dimensions: The tops of so-called 'dynamic fibrils' in
active regions and mottles in quiet Sun show parabolic trajectories as
function of time. RMHD models show that they can be
explained as slow-mode acoustic shocks that propagate only along the
direction of the magnetic field
\citep{2006ApJ...647L..73H,2007ApJ...655..624D,2007ApJ...660L.169R,2013ApJ...776...56R}.
Another class of fibrils is present in network areas and active
regions. They appear as dense rows of elongated dark features that
span the chromosphere between areas with opposite magnetic polarity in
the photosphere. The horizontal size of these fibrils ($\sim30$~Mm) is
much larger than the extent of the optically thick \Halpha\ core
emission above the solar continuum limb
\citep[$< \sim$5~Mm, see, e.g,][]{1966ApJ...143...38S},
and estimates of the typical \Halpha\ formation height of 0.5~Mm to 3~Mm
\citep{1981ApJS...45..635V,2012ApJ...749..136L}
Therefore, in a simplified view where we assume magnetic field lines
to be semi-circular, we do not expect these fibrils to trace the
magnetic field in 3D, because the loop height is much larger than
the \Halpha\ formation height. While field lines in reality will not
be semi-circular, the relation between fibrils and field lines remains
unclear. 

Understanding this relationship is relevant because \Halpha\ is one of
the most-used diagnostics of the chromosphere, so improving our
knowledge of its line formation will help understanding the physics of
the chromosphere. In addition, \Halpha\ is used as a constraint on
magnetic field extrapolations
\citep[e.g.,][]{2008SoPh..247..249W,2011ApJ...739...67J},
which requires understanding the
fibril-field line relation. Finally, transverse oscillations in
fibrils are used in chromospheric seismology.
 
Chromospheric seismology aims to derive properties of the chromosphere through
measuring properties of observed oscillations and interpreting them as
a certain types of waves. Oscillations of fibrils seen in the
\Halpha\ line core have been interpreted as {\it kink waves}:
transverse incompressible oscillations of a tube of high mass density
that is aligned with the magnetic field and is embedded in a medium
with lower mass density. In this view, measurement of the periods $P$,
wave displacement amplitudes $A$ and phase speeds of perturbation of
the \Halpha\ fibrils, and possibly their variation along the length of
individual fibrils, then allows derivation of chromospheric
properties. The period and displacement can be converted to velocity
amplitudes $V$ if one assumes sinusoidal motion. The velocity
amplitude together with an assumed mass density allows computation of
the energy density of the wave. If phase speeds are measured then one
can also derive the strength of the magnetic field and the energy flux
carried by the wave.

In an observation of quiet Sun including some network boundary
\citet{2012ApJ...750...51K} % kuridze et al
measured lifetimes of 42 mottles/fibrils and the period and amplitude of their
transverse oscillations in a time-series of \Halpha\ line-center
images. 
\citet{2013ApJ...768...17M} 
analyzed the same dataset in more detail and derived improved
statistics on the period and displacement of transverse fibril
oscillations, finding an average period of 94~s and an average displacement
of 71~km.

\citet{2013ApJ...779...82K} looked at a different quiet Sun dataset
and presented three fibrils for which apparent phase speeds were
measured. They found evidence of apparent phase speeds increasing from
40 \kms\ to 120 \kms\ along one fibril, and a fibril with signal of
opposite phase speeds of +101 \kms\ and -79 \kms\ respectively,
suggestive of two waves traveling in opposite direction. They also
reported on a fibril oscillation with an apparent phase speed of 350
\kms, which they interpret as a standing wave pattern caused by
oppositely directed waves.

In a recent paper,
  \citet{2014ApJ...784...29M}
derived a velocity power spectrum of transverse \Halpha\ oscillations
in a quiet-Sun region and an active region and compared the power
spectrum with the coronal transverse motion power spectrum derived
from observations with the Coronal Multi-channel Polarimeter
instrument

  \citep{2009ApJ...697.1384T}, %tomczyk and mcintosh
to estimate the frequency-dependent damping of  transverse waves between the
chromosphere and corona.

The above discussion shows that \Halpha\ fibril seismology has the
potential to be a powerful tool for diagnosis of chromospheric
physical properties and the transport of energy by transverse
oscillations from the photosphere into the outer solar
atmosphere. However, the validity of this approach depends critically
on two related assumptions:
\begin{enumerate} \itemsep1pt \parskip0pt \parsep0pt
\item fibrils trace single magnetic field lines at any given instant in time;
\item fibrils trace the same field line during their lifetime.
\end{enumerate}

The fibrils most likely do trace out the horizontal component of the
magnetic field, but as we have argued, it is not known whether fibrils
also follow the vertical component of the field.

For the second assumption there is no observational evidence. We argue
that it not at all clear that this is the case:
\citet{2012ApJ...749..136L} 
show that low \Halpha\ line-core intensity typically corresponds to a
large formation height, and that the line opacity is proportional
to the mass density. Their simulation shows that fibrils correspond to
mountain-ridge like structures of increased mass density. Dark fibrils
are thus most likely structures that are over-dense in the horizontal
direction compared to their surroundings.  Even if a fibril traces out
a single field line at a given instant in time, it is thinkable that
this field line is drained of its mass while a neighbouring field line
gets filled with mass. The \Halpha\ fibril will then appear to move,
while the actual field lines do not.

The main aim of this paper is to investigate the relation between
magnetic field lines in a RMHD simulation of the solar atmosphere and
the fibril-like structures (from now on we call them just fibrils)  in a time series of synthetic
\Halpha\ images computed from the simulation. We use the results of
the investigation to see to what extent the assumptions made in
chromospheric seismology are valid, by analysing the time
evolution of the fibrils and magnetic field lines that thread
through them. Finally we analyse the evolution of the atmosphere along
those fibril-threading field lines to investigate the dynamics of the
simulated chromosphere.

In Section~\ref{sec:simulations} we describe the radiation-MHD
simulation and the subsequent radiative transfer
modelling. Section~\ref{sec:observations} describes the observations
that we use. In Section~\ref{sec:fibrilosc} we treat the fibrils in the synthetic
images as observations to establish that they show the same
oscillatory behaviour as the observations.  Section~\ref{sec:reduct}
describes the reduction of the RMHD simulation data to extract
properties along selected magnetic field lines and in
Section~\ref{sec:analysis} we analyse the relation between the fibrils
and the magnetic field lines. We finish with a discussion and our
conclusions in Section~\ref{sec:discussion}.

% *******************************************************************
\section{Simulations and radiative transfer} \label{sec:simulations}
% *******************************************************************

% *******************************************************************
\begin{figure}  \itemsep1pt \parskip0pt \parsep0pt
  \includegraphics[width=\columnwidth]{\figspath/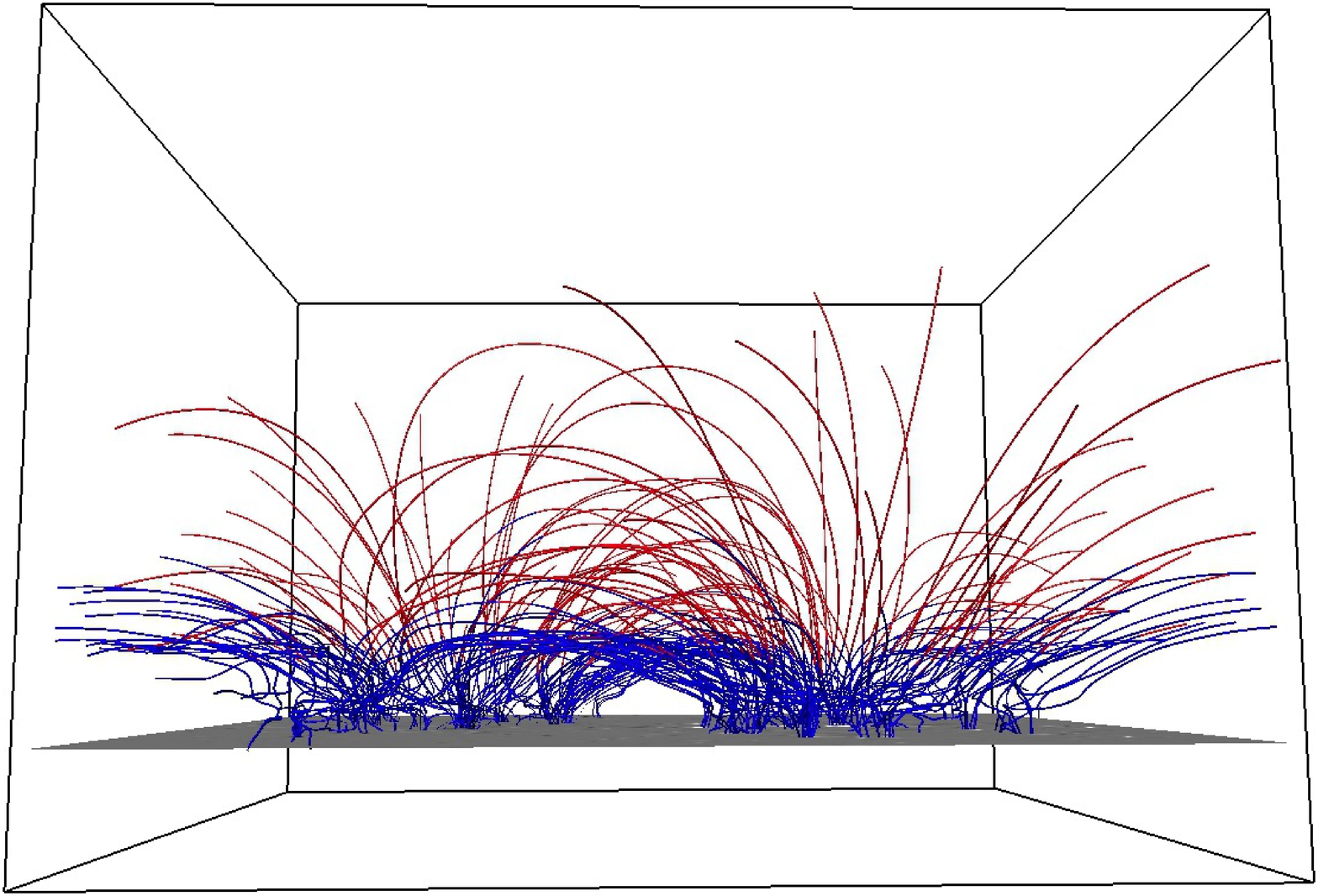}
  \includegraphics[width=\columnwidth]{\figspath/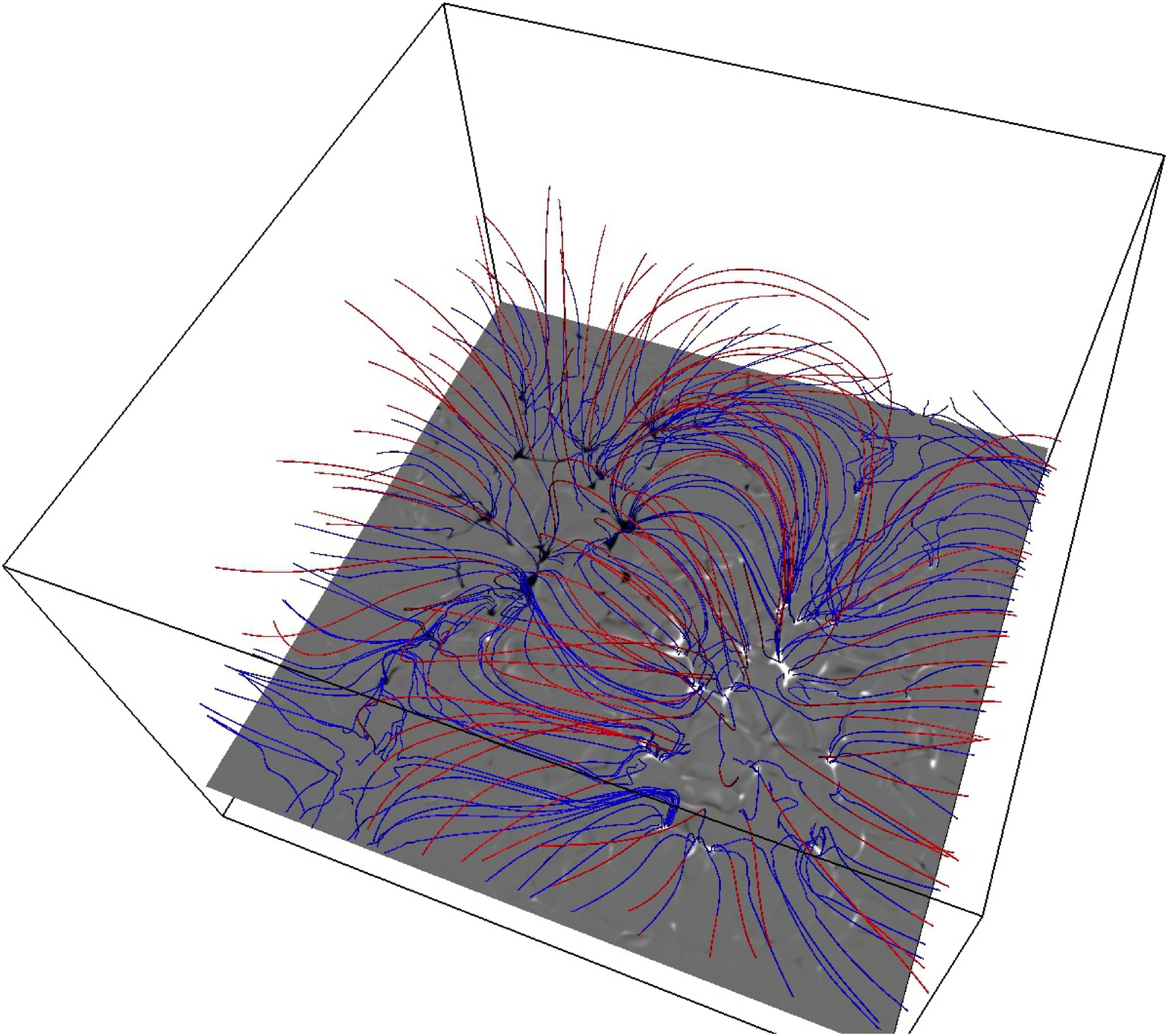}
  \includegraphics[width=\columnwidth]{\figspath/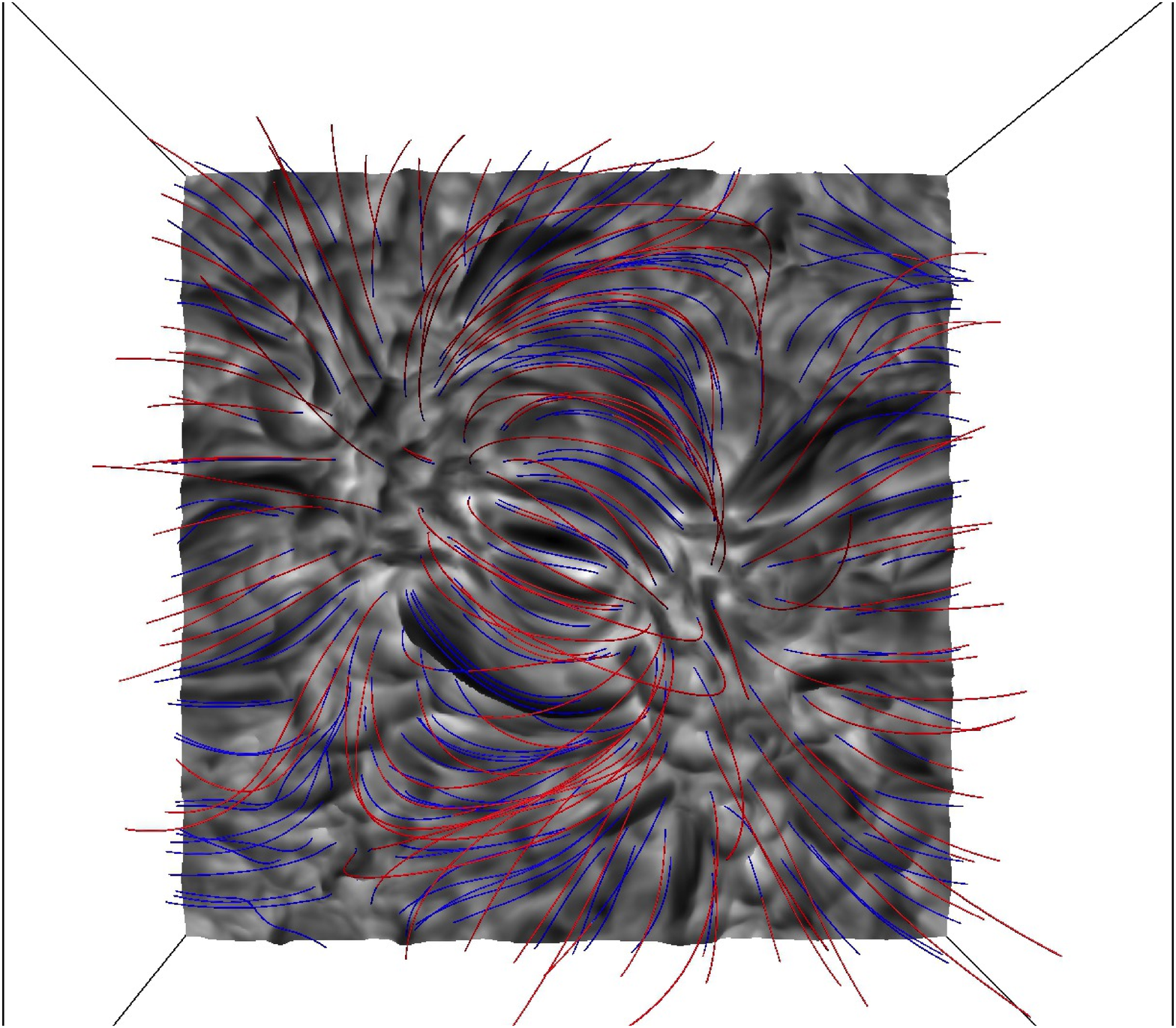}
   \caption{Magnetic field configuration in the RMHD simulation. The
     wire frame outlines the border of the computational domain. Each panel shows
     the same set of magnetic field lines, color coded with the local
     temperature, blue for $T < 20$~kK, red for $T \ge 20$~kK. {\it
       Top and middle panel:} side and top view of the domain. The
     gray-scaled plane displays the vertical magnetic field at
     z=0~Mm. {\it Bottom panel:} The gray-scaled surface is the
     \Halpha\ line-core height of optical depth unity, color coded
     with the corresponding \Halpha\ vertically emergent
     intensity. Only the section of the field lines above the surface
     of optical depth unity is displayed. The visualization was created with
     VAPOR \citep{2007NJPh....9..301C}. 
  \label{fig:snap60}}
\end{figure}
% *******************************************************************

We performed a 3D radiation-MHD simulation of a part the solar
atmosphere using the \Bifrost\ code
\citep{2011A&A...531A.154G}.
This code solves the equations of resistive MHD on a staggered
Cartesian mesh. In addition to the MHD equations the simulation
included optically thick non-LTE radiative transfer in the photosphere
and low chromosphere, parameterized radiative losses and gains in the
upper chromosphere, transition region and corona, thermal conduction
along field lines and an equation of state that takes the
non-equilibrium ionization of hydrogen into account.

The simulation was run on a grid of $504 \times 504 \times 496$ grid
cells, with an extent of $24 \times 24 \times 16.8$ Mm. In the
vertical direction the grid extends from 2.4 Mm below to 14.4 Mm
above average optical depth unity at 500 nm, encompassing the upper
convection zone, photosphere, chromosphere and lower corona. The $x$
and $y$-axes are equidistant with a grid spacing of 48 km. The
$z$-axis is non-equidistant. It has a grid spacing of 19 km between
$z=-1$ and $z=5$ Mm, while the spacing increases towards the upper
and lower boundaries to a maximum of 98 km at the coronal
boundary. The simulation has periodic boundary conditions in the two
horizontal directions and open upper and lower boundaries.

The magnetic field configuration is mainly bipolar, chosen such that
it leads to a small network-like configuration. It was created by
specifying the magnetic field at the bottom boundary and using a
potential field extrapolation to compute the field throughout the
computational domain. The magnetic field was inserted into a
relaxed hydrodynamical simulation and was then allowed to evolve
freely. The simulation was subsequently run for 3000 s of solar time using
LTE ionization, and then the non-equilibrium hydrogen ionization was
switched on. Afterwards the simulation was run for 2240 s of solar time. We
use a dataset of 1200 s of solar time (121 snapshots taken at 10 s
intervals) starting 830 s after non-equilibrium ionization was
switched on. The first snapshot of our time series is the same as was
used in
\citet{2012ApJ...749..136L,2012ApJ...758L..43S,%
       2013ApJ...772...89L,2013ApJ...772...90L}
and
\citet{2013ApJ...778..143P}.
The snapshots are publicly available for
download\footnote{\url{http://sdc.uio.no/search/simulations}}.
The whole simulation is described in more detail in Carlsson et al. (2015).

The synthetic \Halpha\ imagery was computed following 
\citet{2012ApJ...749..136L}.
In short, we use the radiative transfer code \multitd\
\citep{2009ASPC..415...87L}
to solve the non-LTE radiative transfer problem in 3D geometry for a
5-level-plus-continuum hydrogen atom. Owing to the strong scattering
in the Lyman and Balmer lines it is essential to include 3D
effects. The Lyman-$\alpha$ and Lyman-$\beta$ lines are treated with
Doppler absorption profiles only to approximate PRD effects. In order
to save computation time we down-sampled the \Bifrost\ snapshots to
half resolution in the horizontal direction, and solved the radiative
transfer only every second snapshot, i.e, 61 snapshots at 20~s cadence with a grid
size of $252 \times 252 \times 496$ cells. This is a slight
improvement over \citet{2012ApJ...749..136L}, where the atmosphere was
also down-sampled in the vertical direction. We performed a test
computation at the full resolution of the RMHD simulation, and found
that the down-sampling reduces image sharpness but otherwise does not
affect our results.

In Figure~\ref{fig:snap60} we show a characterisation of the magnetic
field in the simulation at $t=600$~s in our time series. The magnetic
field exhibits a large-scale bipolar structure in the photosphere; as
a consequence the chromosphere and corona are pervaded by loop-like
structures. The majority of the magnetic field closes in the
chromosphere: the average unsigned field at $z=0$ is 29.3~G, at
z=0.8~Mm it reaches only 50\% of this value and at z=3.3~Mm 25\%. The
bottom panel of Figure~\ref{fig:snap60} shows an iso-surface of the
\Halpha\ line-core height of optical depth unity, colored with the
corresponding \Halpha\ intensity, illustrating that the fibrils
roughly align with the horizontal component of the magnetic field.

% *******************************************************************
\section{Observations}  \label{sec:observations}
% *******************************************************************

% *******************************************************************
\begin{figure*}
  \includegraphics[width=17cm]{\figspath/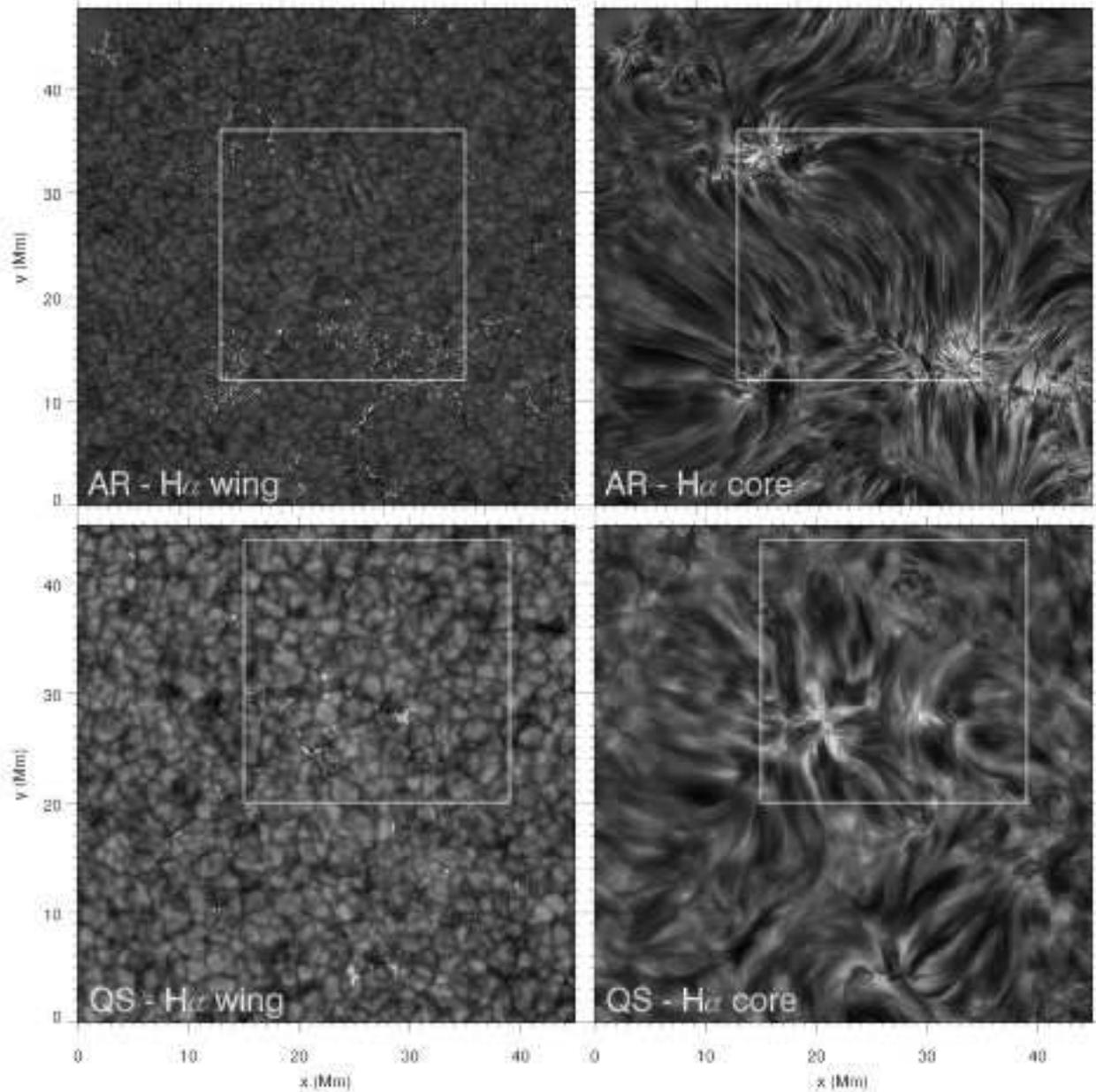}
  \caption{Example \Halpha\ images of the active region dataset taken
    on June 15, 2008 (top) and the quiet Sun dataset taken on May 5,
    2011 (bottom). The left-hand column shows images in the red wing
    of the line, the right-hand column the profile-minimum intensity.
    The subregions used in Figure~\ref{fig:obs_sim_fibrils} are
    indicated by white boxes. }
  \label{fig:AR_QS_ims}
\end{figure*}
%*******************************************************************

In Section~\ref{sec:fibrilosc} we compare the synthetic
\Halpha\ imagery with observations. These observations were obtained
with the CRisp Imaging SpectroPolarimeter
\citep[CRISP,][]{2008ApJ...689L..69S} % scharmer et al. CRISP
at the Swedish 1-m Solar Telescope
\citep[SST,][]{2003SPIE.4853..341S} % scharmer et al. SST
on La Palma. 

We use a dataset of the remains of active region AR10998 taken on June
15, 2008 at 7:53 UT, with a duration of 27 minutes at 6.8 s
cadence. The data were taken close to disk-center, at $\mu=0.98$. 
The \Halpha\ line was sampled at 25 positions, equidistant 
with 0.1~\AA\ steps between
$-1.6$~\AA\ and $+0.8$~\AA\ from line center. We refer to this dataset
as Active Region (AR).

A second dataset was acquired on May 5, 2011 at 7:38 UT.  The
\Halpha\ spectral line was sampled with 35 line positions, from
$-$2.064~\AA\ to $+$1.290~\AA, and equidistant 86~m\AA\ stepping
between $\pm$1.290~\AA. The time to complete a full line scan was 8~s.
The target area was a quiet 60$\times$60\arcsec region at
$\mu=1.0$. We refer to this dataset as Quiet Sun (QS).

Both datasets were captured during excellent seeing conditions, and
the image quality
was improved by the
adaptive optics system of the SST
\citep{2003SPIE.4853..370S} % Scharmer et al: SST AO
and post-processing using Multi-Object Multi-Frame Blind Deconvolution
\citep[MOMFBD,][]{2005SoPh..228..191V}
image restoration. For the data reduction we followed early versions
of parts of the CRISPRED data processing pipeline
\citep{2014arXiv1406.0202D}. % Jaime et al CRISPRED
Figure~\ref{fig:AR_QS_ims} shows example images of the AR and QS
datasets.

% *******************************************************************
\section{Comparison of observed and simulated \Halpha\ fibrils}  \label{sec:fibrilosc}
% *******************************************************************

% *******************************************************************
\subsection{\Halpha\ as velocity diagnostic of fibrils}
% *******************************************************************

% *******************************************************************
\begin{figure}
  \includegraphics[width=\columnwidth]{\figspath/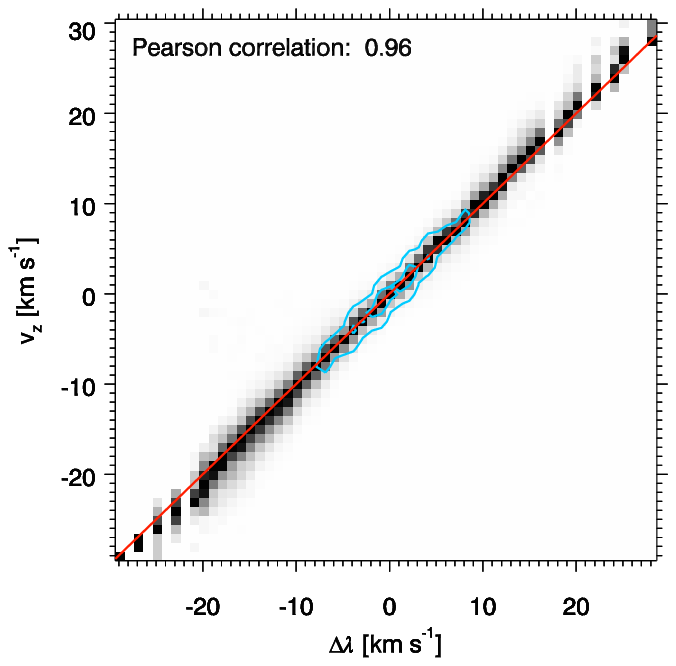}
  \caption{Scaled joint probability density functions of the Doppler
    shift of the \Halpha\ profile minimum (positive is redshift) and
    the vertical velocity at optical depth unity at that Doppler shift
    (positive is down flow). Each column in the panels is scaled to
    maximum contrast to increase visibility. The inner blue contour
    includes 50\% of all pixels, the outer contour 90\%. The red line
    denotes the line $v_z = \Delta \lambda$.}
  \label{fig:dv_vz_corr}
\end{figure}
% *******************************************************************

Our SST observations include a
sampling of the line profile and we can thus use them to compare
Doppler-shifts in the observed and simulated datasets. Because fibrils
are most clearly visible in line-core images, we use the Doppler-shift
of the profile minimum as their velocity diagnostic.

\citet{2013ApJ...772...90L} showed that the Dopplershift of the
central line-profile minima of the \MgII\ h\&k lines exhibit
remarkable correlation with the vertical velocity at the height of
optical depth unity. They are thus excellent velocity diagnostics. The
high correlation for the \MgII\ lines is caused by the
small ($\simeq 2.5$~\kms) Doppler width of the absorption
profile and the steep increase of optical depth with geometric depth
causing only a small height range to be sampled. The \Halpha\ line has
five times larger Doppler width. Numerical simulations indicate that its
contribution function to intensity is often non-zero over a height
range of 1 Mm or more
\citep{2012ApJ...749..136L}.
We thus expect \Halpha\ to perform somewhat worse as a velocity
diagnostic than \MgII\ h\&k.

We computed the correlation between the Doppler shift of the
line-profile minimum and the vertical velocity at optical depth unity
at the wavelength of the profile minimum based on all pixels in our synthetic
time-series. 

The results are displayed in
Figure~\ref{fig:dv_vz_corr}. The correlation is indeed not as good as
for \MgII~h\&k, but its Pearson correlation coefficient is still very
high at 0.96. We conclude therefore that the \Halpha\ line is a good
velocity diagnostic for those layers that are sampled by the profile
minimum. Test with the observational datasets show that traditional
Dopplergrams created by subtracting and normalizing intensity images
at equal wavelength separation from line center are visually very
similar to the Doppler-shift of the profile minimum. The advantages of
using profile-minimum Dopplershift maps is that it provides the
vertical velocity in absolute units, and that the profile-minimum
intensity provides a direct visualization of the structures -- fibrils
in this paper -- for which the velocity is measured.

% *******************************************************************
\subsection{Comparison of the appearance of synthetic and observed \Halpha\ fibrils}
\label{sec:compsyntobs}
% *******************************************************************

% *******************************************************************
\begin{figure*} 
  \includegraphics[width=17cm]{\figspath/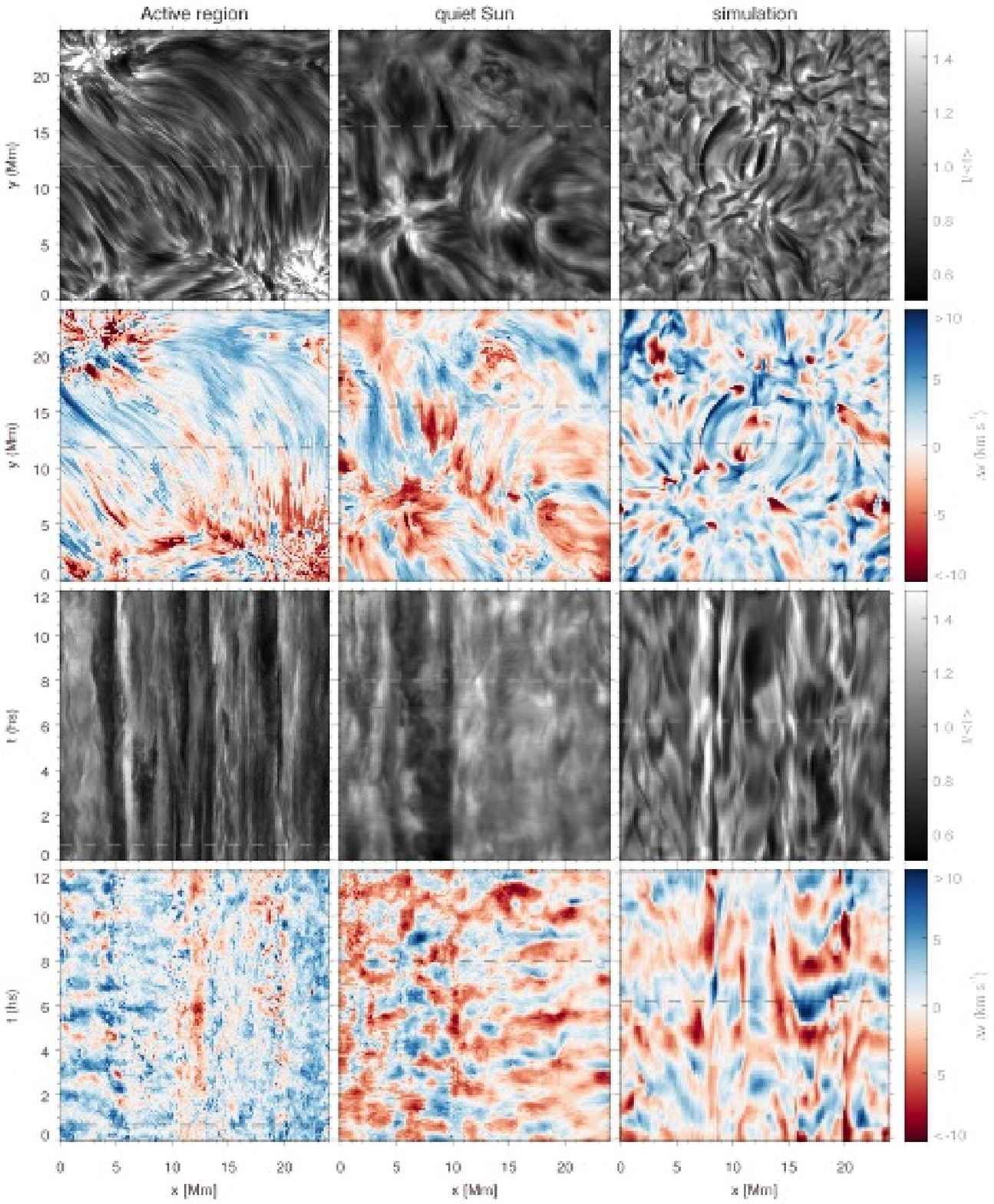}
  \caption{Comparison of images and space-time diagrams of fibrils in
    an active region (left), quiet Sun (middle) and the simulation
    (right). Rows, from top to bottom display \Halpha\ profile-minimum
    intensity images, \Halpha\ line-core Doppler-shift images,
    intensity space-time diagrams, and Doppler-shift space-time
    diagrams. The dashed lines specify the location of the space-time
    cut in the images and vice versa. Blue color in the Doppler-shift
    panels corresponds to upflow. Time is given in units of
    hecto-seconds ($10^2$s ) in this and all following figures.}
  \label{fig:obs_sim_fibrils}
\end{figure*}
% *******************************************************************

The RMHD simulation incorporates many of the physical processes that
act in the solar chromosphere and indeed shows fibrilar structure
reminiscent of observations.  However, the separation of the opposite
polarity patches in the photosphere is only $\sim 10$~Mm, considerably
smaller than the typical size of a supergranule
\citep[$\sim 30$~Mm, e.g.][]{1964ApJ...140.1120S}, %simon&leighton Velocity Fields in the Solar Atmosphere. III. 
and the maximum length of the simulated fibrils is thus shorter than
possible on the sun. The finite resolution of the simulation (48~km
horizontal, 19~km vertical) does not allow horizontal structures
smaller than $\sim 100$~km to form. It is therefore worthwhile to
analyze in more detail how the simulated fibrils appear and behave in
comparison with observed ones.

In Figure~\ref{fig:obs_sim_fibrils} we compare fibrils in the AR and
QS observations and the simulation. The panels showing the
observations display the same spatial and temporal extent as the
simulation to allow for fair comparison.

The AR area has a cluster of photospheric magnetic elements in the
upper-left and lower-right corners of the field-of-view (FOV). Between
these elements a carpet of long and finely structured fibrils is
present. The QS image displays a few broad and short
dark fibrils
\citep[also called mottles, see ][and references therein]{2012SSRv..169..181T}
that extend towards to top of the image. The corresponding \Halpha\ wing
image shows a small group of photospheric magnetic elements in the
lower part of the image (see Figure~\ref{fig:AR_QS_ims}).

The simulated fibrils appear as a cross between the AR and QS ones,
they typically are not as wide as in the QS, but also not as long,
slender and densely packed as in the AR area. The simulated intensity
contrast is similar to the observed one.

The second row shows images of the Dopper-shift of the line
core. The overall structuring is similar to the intensity images. Both
the AR and QS images show upward and downward moving fibrils, but the
AR panel shows Dopper-shift variation perpendicular to the magnetic
field on smaller spatial scales. 

The two bottom rows compare time-distance diagrams of the line-core
intensity and Doppler-shift. The line-core panels (third row) show the
characteristic swaying motions of the fibrils. The swaying fibril
tracks are ubiquitously present in all three datasets. The AR dataset
typically shows tracks that remain visible during a longer time than
in the QS panel, and the simulation again shows characteristics
somewhere in between the AR and QS data.

In the bottom row we show the time evolution of the profile-minimum Doppler
shift along the same 
cut across the fibrils. All three panels show
elongated structure only at the location of the very darkest
fibrils. The rest of the panel is dominated by oscillations with
periods of 3 to 5 min and a coherency length along the slit of $\sim
4$ Mm in the AR and $\sim 10$ Mm in the QS observations.  These
coherent oscillations are most likely the result of buffeting from
below by internetwork acoustic shocks excited by $p$-modes and
granular buffeting
\citep{2008SoPh..251..533R}. %Rutten & van veelen
The simulation displays an oscillation with a typical period of 4 min
that is coherent along the whole 24~Mm cut. This is the signature of
shocks driven by global radial box oscillations -- the simulation
counterpart of $p$-modes -- excited by the convection
\citep{2001ApJ...546..585S}. % stein and nordlund

The AR panel shows smaller Doppler-shift amplitude ($\pm 7$ \kms, RMS
of 1.7 \kms) than the QS (amplitude -8 to +14 \kms, RMS of 2.5 \kms)
and simulated (amplitude -12 to +11 \kms, RMS of 2.9 \kms) data.

It is unclear what is causing the difference in pattern between the
intensity and Doppler shift time-distance diagrams.  It is possible
that many of the fibrils have an optical thickness of close to unity,
so that the Doppler-shift of the profile minimum is partially
determined by the sub-canopy domain, which is full of acoustic waves
and shocks. Another possibility is that we see the actual upward
and downward motion of the fibrils themselves, and that this is
residual motion caused by the reflection of sub-canopy waves at the
$\beta=1$ surface, as demonstrated in the simulations of
\citet{2002ApJ...564..508R}. %rosenthal et al
The reflection of the acoustic waves can have a significant signal
into layers where $\beta<0.1$, where fibrils are presumably
located. This scenario is supported by the fact that the Doppler-shift
amplitude in the AR dataset, where \Halpha\ forms much higher above
the $\beta=1$ surface, are smaller than in the QS, where
\Halpha\ samples closer to the $\beta=1$ surface. Of course another
mechanism altogether remains possible too.

Focusing back on our present purpose: we have established that the
simulation qualitatively reproduces many of the features of the
appearance of fibrils in network and quiet Sun, in both intensity and
Doppler shift. The fibril density and width in the simulations are
intermediate between AR and QS, the simulated Dopplershifts are closer
to the QS properties than to the properties of the AR.

% *******************************************************************
\subsection{Transverse oscillations of synthetic fibrils}
% *******************************************************************

%*******************************************************************
\begin{figure} 
  \includegraphics[width=8.8cm]{\figspath/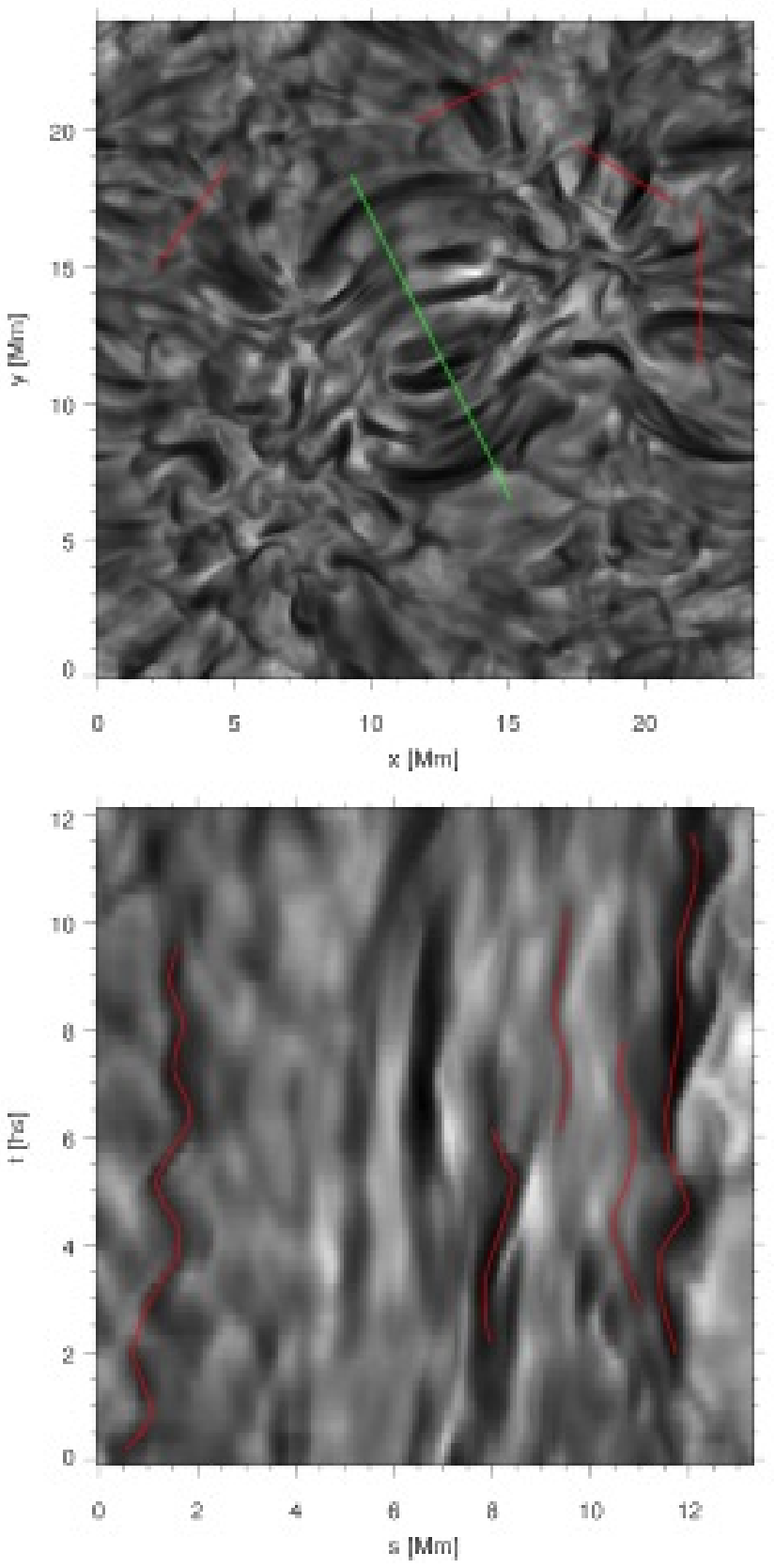}
  \caption{Top panel: synthetic \Halpha\ line-core image at t=600~s.
    The location of the time-distance diagrams used to measure fibril
    oscillation motions are overplotted. Bottom panel: time-distance
    diagram along the green cut in the upper panel, with fibril tracks
    overplotted in red.}
  \label{fig:cuts}
\end{figure}
%*******************************************************************

%*******************************************************************
\begin{figure*} 
  \includegraphics[width=17cm]{\figspath/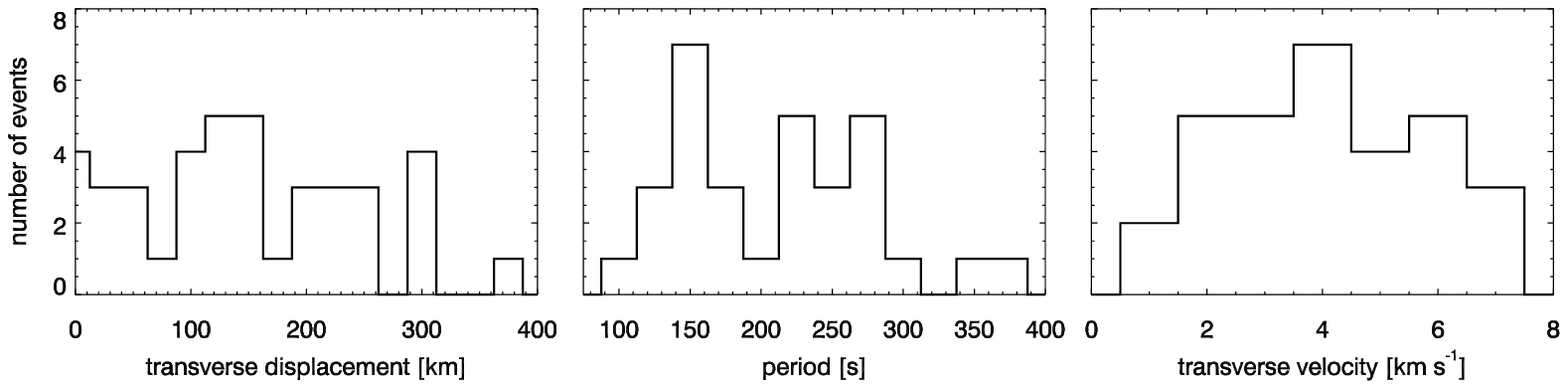}
  \caption{Histogram of the properties of the oscillations of the
    simulated fibrils. Left: maximum transverse displacement; middle:
    period; right: maximum transverse velocity.}
  \label{fig:fibril_histo}
\end{figure*}
%*******************************************************************

We now turn our attention to the periods, displacement and velocity
amplitudes of the synthetic fibrils. After manual inspection of the
data we placed five cuts through the data that showed clear
oscillations of dark fibrils (see Figure~\ref{fig:cuts}). There are
many more fibrils that appear to sway when playing the image sequence
as a movie, but whose time-distance diagram does
not show an unambiguous swaying
track. After selecting the cuts with clear swaying we manually traced
the swaying fibril tracks. We also examined automated detection
algorithms after unsharp masking of the data to bring out contrast,
but found that this leads to many false detections. These are in
particular connected with the bright-dark pattern of the box
oscillations, that after unsharp masking often give rise to sinusoidal
tracks that are not associated with dark fibrils.

In the cuts we found 13 fibrils that could be unambiguously followed
for one period or more. Because we found that the fibrils show
oscillatory behavior, but rarely with constant amplitude and period,
we also measured periods and amplitudes by hand. For each space-time
track we removed a linear trend. Then we measured the space and time
coordinate of each extremum ($t_i,x_i$). The time difference between
two consecutive maxima or minima gives one measurement of a period
$P$: $P_i=t_{i+1}-t_{i-1}$. The spatial difference between two
consecutive extrema yields a measurement of the transverse
displacement $d$: $d_i=0.5(x_{i+1}-x_i)$. Finally we computed the
velocity amplitude $v$ as $v_i=\pi (d_{i-1}+d_{i}) / P_i$. In total we
measured 40 displacements, 31 periods and 31 velocity amplitudes.

The results of our measurements are shown in
Figure~\ref{fig:fibril_histo}. The average (standard deviation) of the
displacement is 158~km (98 km), for the period it is 221~s (70~s) and
for the velocity amplitude it is 4.5~\kms\ (1.8~\kms).
\citet{2013ApJ...768...17M} % morton et al
measure smaller average displacements and periods, 71~km and 94~s.  We
believe this to be at least partially caused by an event selection
bias caused by the higher spatial and temporal resolution of their
observations and their automated versus our manual event
selection. The observations used by
\citeauthor{2013ApJ...768...17M}
have a spatial pixel size of 50~km and a temporal resolution of 7.7~s,
compared to 96~km and 20~s for the simulations. The observations thus
in principle allow detection of oscillations at half the amplitude and
2.6 times smaller period. The automated detection algorithm used by
\citet{2013ApJ...768...17M}
also picks up weak signals, while we focus
on the oscillations of the most prominent dark fibrils. Another possible
source of the difference is the resolution of the simulation, that
might not support oscillations below a certain displacement amplitude.

The average velocity amplitude that we find is the same as
\citet{2013ApJ...768...17M}, 
but the synthetic distribution lacks the tail of high amplitude events
detected in the observations.

In summary, we find that the swaying fibrils in the synthetic
\Halpha\ imagery have periods, maximum displacements and velocity
amplitudes that are consistent with those observed. The lack of
short-period and low-displacement events are possibly caused by finite
spatial resolution of the simulation, the spatial resolution and
cadence of the synthetic images and a selection bias.

% *******************************************************************
\subsection{Phase speeds} \label{sec:phase_speeds}
% *******************************************************************

%  ---------------------------------------- --------------------------
\begin{table}
\caption{Approximate phase speeds in synthetic fibrils (\kms)}
\label{table:phase_speeds}
\centering

\begin{tabular}{ccc}
\hline\hline 
 54  & 18 & 49 \\
 33  & 60 & 56 \\ 
 320 & 27 & 16 \\
\hline
\end{tabular} 
\end{table}
%  ---------------------------------------- --------------------------

% *******************************************************************
\begin{figure} 
  \includegraphics[width=8.8cm]{\figspath/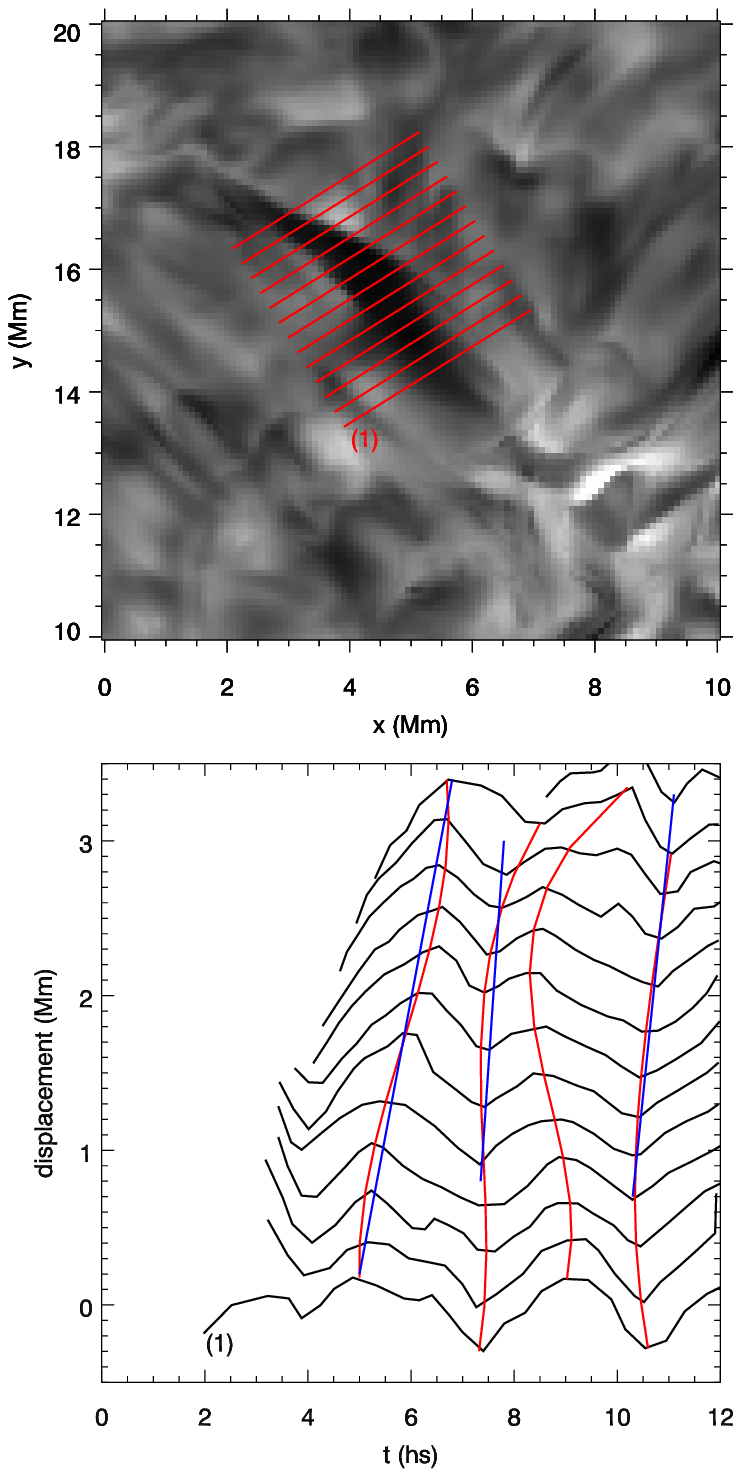}
  \caption{Phase speeds in a simulated fibril. Top panel:
    \Halpha\ image of the dark fibril at $t=600$~s, with the
    cross-cuts used in the bottom panel overplotted in red. Bottom
    panel: The black curves show the displacement of the fibril along
    each of the cross cuts in the top panel as function of time, with
    a linear trend removed. The track labelled (1) corresponds to the
    cut with the same label in the upper panel.  The other tracks have
    been shifted vertically by the distance between the corresponding
    cross-cut and the cut labeled (1) in the upper panel. Blue lines:
    linear fits to selected extrema. Red curves: third-order
    polynomial fits to the strongest extrema.}
  \label{fig:sim_phase6}
\end{figure}
% *******************************************************************

% *******************************************************************
\begin{figure} 
  \includegraphics[width=8.8cm]{\figspath/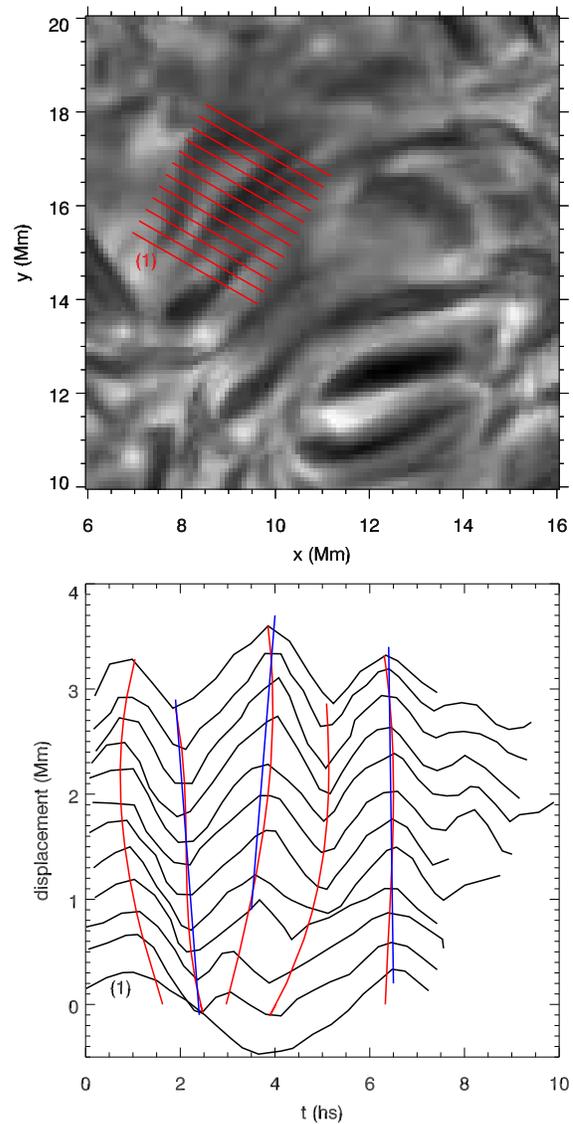}
  \caption{Same as Figure~\ref{fig:sim_phase6}, but for a another fibril.}
  \label{fig:sim_phase8}
\end{figure}
% *******************************************************************

In our synthetic \Halpha\ time series we found a few instances of
strong fibrils for which it was possible to measure phase speeds. We
placed a number of cross-cuts intersecting the fibrils at different
places along their length and measured the $st$-track along each
cut. We then subtracted a linear fit from the tracks to remove
systematic drift. In Figures~\ref{fig:sim_phase6}
and~\ref{fig:sim_phase8} we show two strong, long-lived fibrils and
their swaying along the cross-cuts. The phase speeds of the extrema
are typically not constant, but show acceleration, deceleration and
sometimes even a reversal of direction.  The fibril in
Figures~\ref{fig:sim_phase8} shows clear indication of a minimum
traveling in one direction along the fibril, and the following maximum
in the other direction. The complexity of these phase diagrams is
suggestive of an interference pattern caused by waves traveling in
opposite directions. For those instances where the phase speed was
nearly constant over most of the length of the fibrils we show the
measured speeds in Table~\ref{table:phase_speeds}, they range from 16
\kms\ to 320 \kms. This is consistent with
\citet{2013ApJ...779...82K}, % kuridze et al
who report phase speeds of 79~\kms, 101~\kms\ and 360~\kms.

We conclude that the simulation reproduces the observed properties
sufficiently well to assume that the simulation contains the essential
physics driving transverse oscillations of fibrils.  In
Section~\ref{sec:reduct} we take a closer look at the MHD structures
and their dynamics that give rise to the synthetic \Halpha\ fibrils.

% *******************************************************************
\section{Reduction of the RMHD data} \label{sec:reduct}
% *******************************************************************

We chose to reduce the data using a magnetic field line based
approach. The rationale for this approach is that fibrils form in the
chromosphere in a low plasma $\beta$ environment. Both slow waves and
Alv\'en waves therefore propagate parallel to the field lines, and
analysis of the data along field lines then provides a natural way to
bring out those waves.

\begin{figure}  \itemsep1pt \parskip0pt \parsep0pt
  \includegraphics[width=\columnwidth]{\figspath/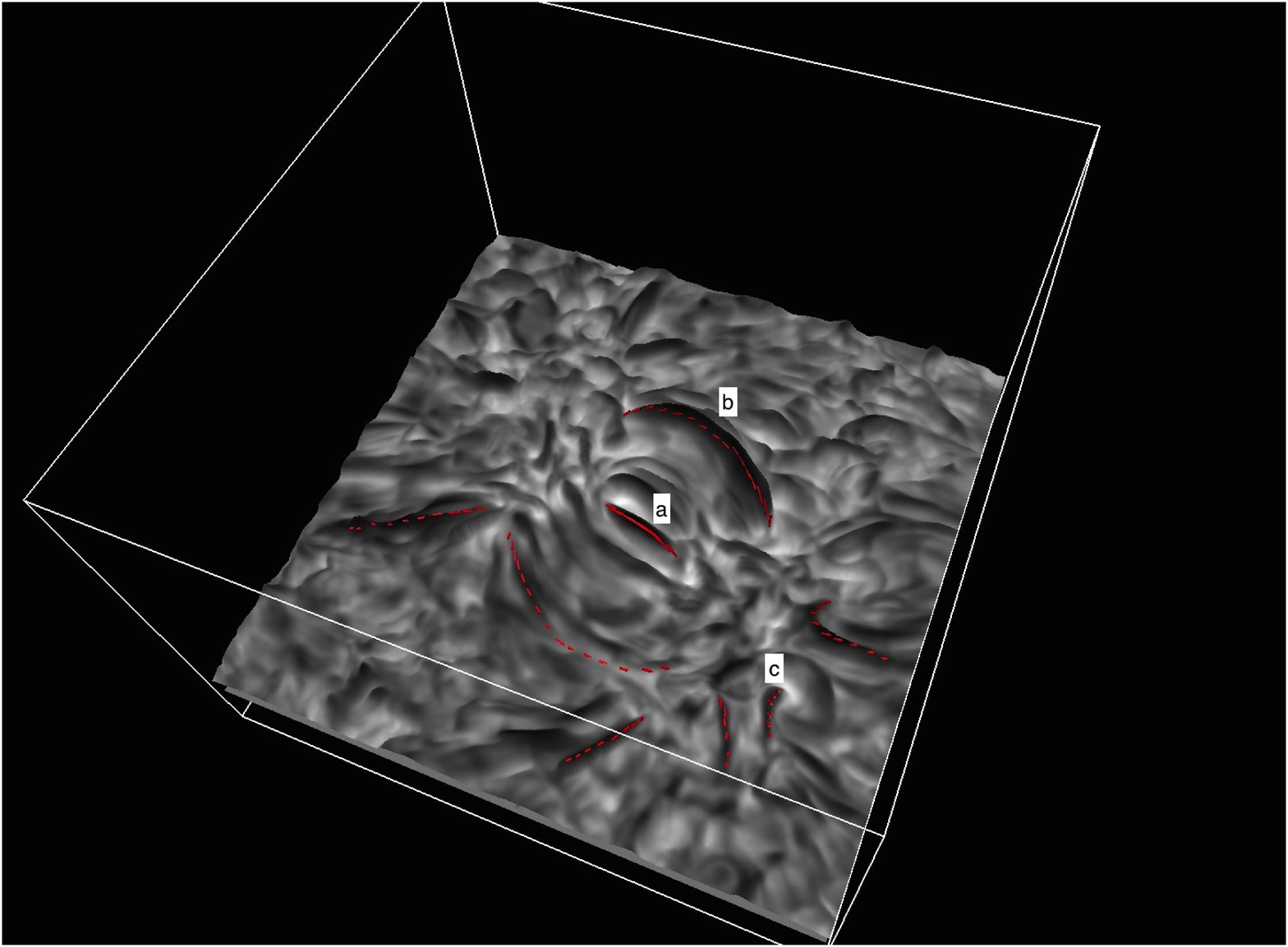}
  \includegraphics[width=\columnwidth]{\figspath/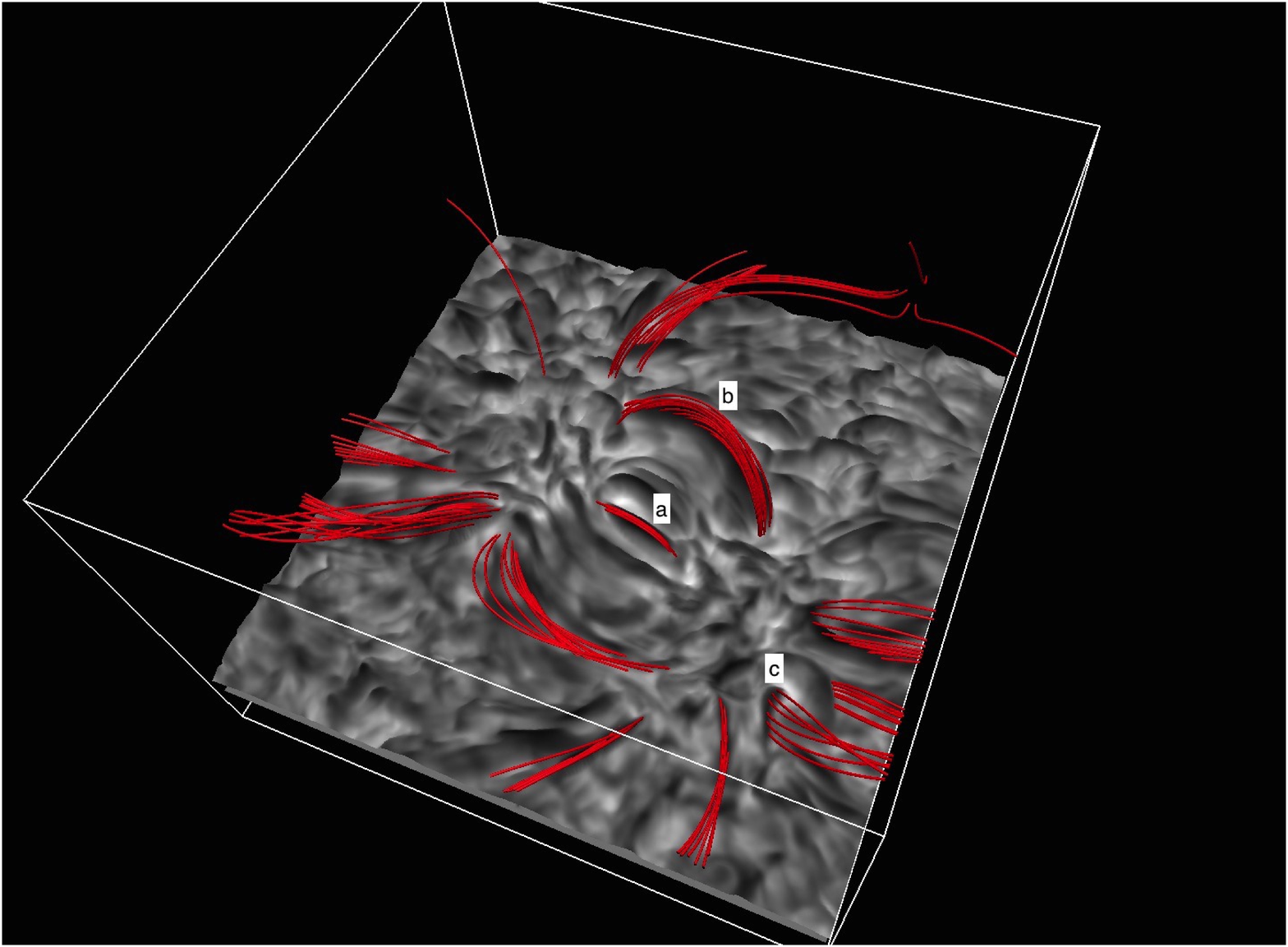}
  \includegraphics[width=\columnwidth]{\figspath/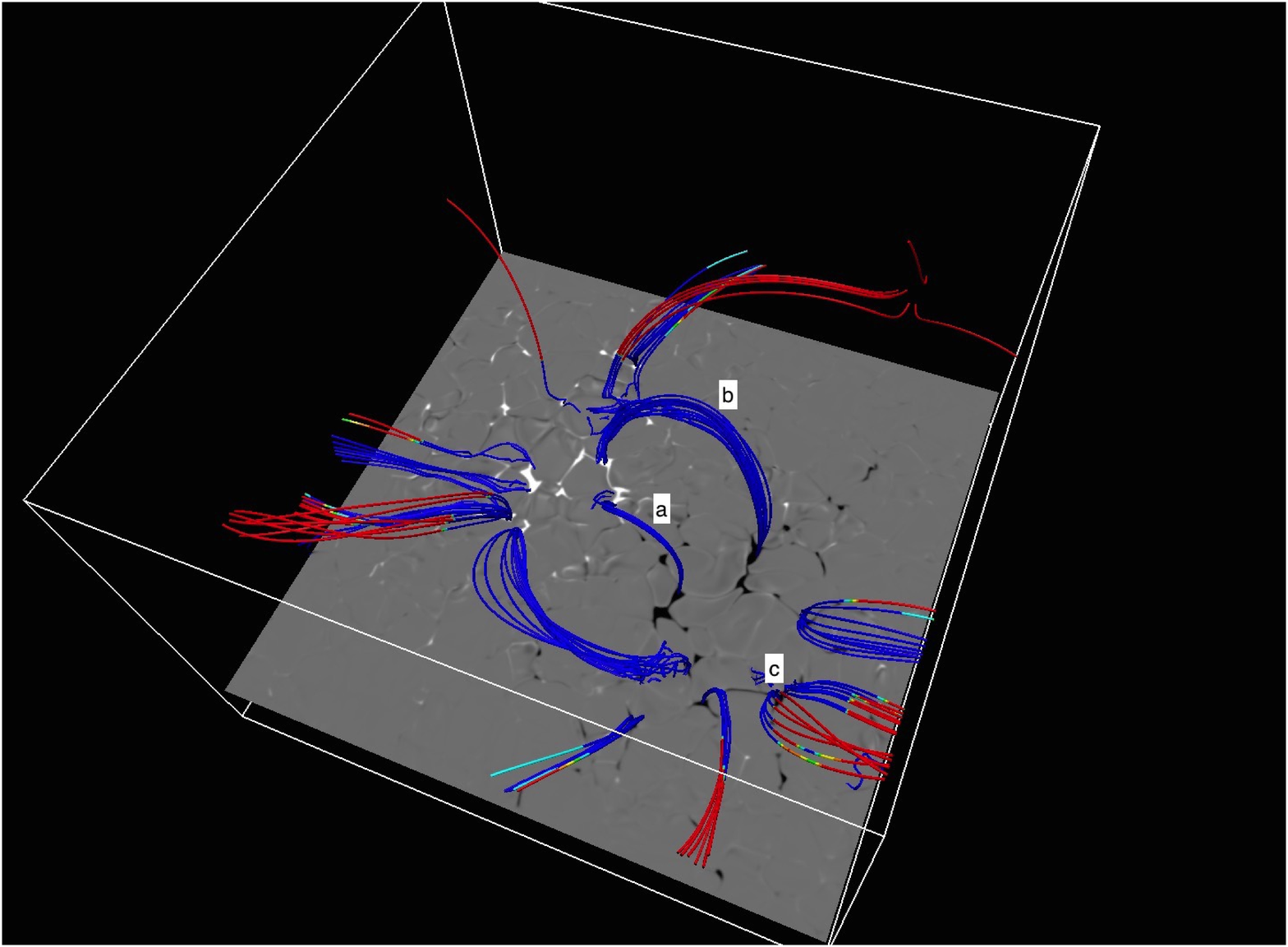}
   \caption{Three-dimensional visualisation of selected field
     lines that thread through \Halpha\ fibrils at $t=600$ s in the
     simulation. Top: Surface of optical depth unity, colour coded with
     the emergent \Halpha\ intensity, with red line segments
     indicating the field line seed points and the local direction of
     the magnetic field.  Middle: Same as top panel, but now showing
     the part of the field lines that extends above the optical depth
     unity surface. Bottom: The same field lines, but now shown on top
     of the vertical magnetic field at $z=0$ Mm. The field lines are
     colour coded with temperature: dark blue for $T<17$~kK; red for
     $T>100$~kK; other colours indicate intermediate values. The labels 
     a, b and c indicate fibrils and associated flux bundles shown in 
     Figure~\ref{fig:fibril_class}.
  \label{fig:flines}}
\end{figure}

We planted 80 seed points in the simulation at $t=600$~s
along a number of fibrils in the corresponding \Halpha\ image. The $x$
and $y$ coordinates of the seed points thus follow the fibrils, and
are spaced about 0.5~Mm apart. As $z$-coordinate of the points we
chose the height of optical depth unity for the profile minimum. From
each seed point we traced a magnetic field line. In order to trace the
same field line over time we advected the field line apex both forward
and backward in time, used the new location to trace a new field line
and repeated this until we obtained the time evolution of the entire
field line.  In Figure~\ref{fig:flines} we show the location
of the seed points and the field lines traced from them.

% *******************************************************************
% *******************************************************************

% *******************************************************************
\begin{figure} 
  \includegraphics[width=8.8cm]{\figspath/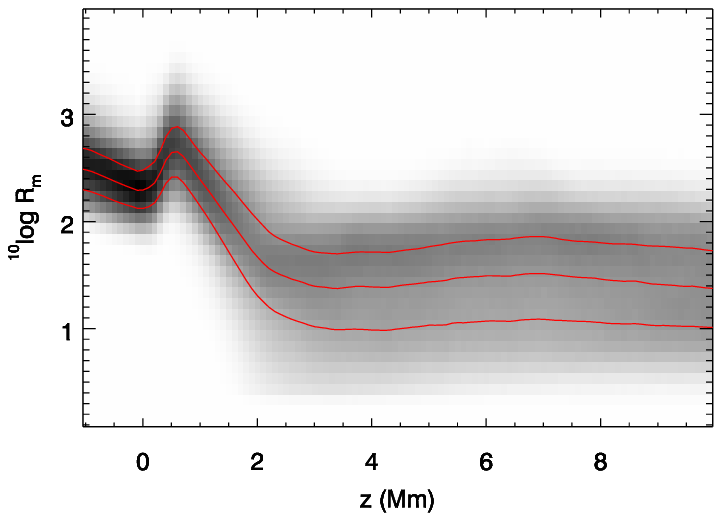}
  \caption{Distribution of the magnetic Reynolds number as function of
    height at t=600~s in the Bifrost simulation.The red curves
    indicate the 25th, 50th and 75th percentile.}
  \label{fig:jpdf_z_rm}
\end{figure}
% *******************************************************************

The validity of this approach is limited by the finite time resolution
of our snapshot series (10~s cadence), which limits the ability to
accurately follow test particles in time, and the magnetic Reynolds number
$R_\mathrm{m}$, which determines how well the assumption of frozen-in
magnetic flux holds. Bifrost employs a space and time variable,
non-isotropic magnetic diffusion coefficient
$\vect{\eta}(\vect{r},t)$, so we cannot give a single value for
$R_\mathrm{m} =(LU)/\eta$. Instead we show the joint probability
density distribution in Figure~\ref{fig:jpdf_z_rm}, assuming a typical
length scale $L$ of 1~Mm and a typical velocity $U$ of 10~\kms. The
Reynolds number is always larger than unity.  Even at heights where
it is smallest, 75\% of the grid cells has $R_\mathrm{m}>10$ and 50\%
has $R_\mathrm{m}>30$. The frozen-in condition thus holds for a large
fraction of the domain, but we also expect locations where diffusion
is important. In Section~\ref{sec:analysis} we show that many field
lines that we analyse remain anchored in the same photospheric field
concentrations for the 1200~s duration that we consider, and evolve
smoothly in time, indicating sufficient time resolution and validity
of the frozen-in assumption.

However, there are also instances where the field lines evolve in
jumps, indicating insufficient time resolution,strong diffusion
or reconnection. In this paper we focus on wave dynamics and we
simply ignore such field lines. We note however that this approach,
improved by proper tracking of the advecting velocity field with test
particles, will yield interesting insights also in case of strong
diffusion and reconnection.

As we are interested in the velocity transverse to the magnetic field
we need to project the velocity onto appropriate axes. A convenient
choice is the Frenet-Serret frame
\citep{1851S,1852F}.

A magnetic field line can be defined as curve $\vect{r}(s)$ through
three-dimensional space as function of the arc length $s$. The curve
is defined by
\be 
\frac{\dd \vect{r}}{\dd s} = \frac{\vect{B}}{| \vect{B} |} = \vect{b},
\ee
given a seed point $\vect{r}_0(s_0)$, where we introduced
$\vect{b}$ as the unit vector pointing in the direction of
$\vect{B}$. The Frenet-Serret frame is an orthogonal coordinate system
at each point along the curve with unit vectors $\vect{T}$,
$\vect{N}$ and $\vect{P}$ given by
\bea
\vect{T} & = & \frac{\dd \vect{r}}{\dd s}  = \vect{b}, \\
\vect{N} & = & \frac{\dd \vect{T}}{\dd s} \left/ {\left| \frac{\dd \vect{T}}{\dd s} \right| }\right. , \\  
\vect{P} & = & \vect{T} \times \vect{N}.
\eea
As an example one can consider a semicircular magnetic field line with
radius $R$ in a plane perpendicular to the photosphere, the latter we take
to lie in the $xy$ plane at $z=0$:
\be
\vect{r}(s) =  \left\{
\begin{array}{ll}
R \cos (s/R) \,\hat{x} \\
0\, \hat{y}\\
R \sin (s/R)\, \hat{z}
 \end{array} 
\right. 
\ee
The $\vect{N}$ vector along each point of the field line then points 
towards the center of the circle at the origin and $\vect{P}$ is
the unit vector in the $y$-direction. If one would observe this field
line at solar disk center and observe transverse oscillations, then
these oscillations would be along $\vect{P}$. For field lines with
more complex 3D geometry, as we have in our simulation, we generally
do not expect that transverse waves that are initially excited only
along either $\vect{N}$ or $\vect{P}$ will remain like that. For our
analysis this will not pose problems: Models of
transverse oscillations along curved loops developed by
\citet{2009SSRv..149..299V}
indicate that the eigenfrequencies, and thus phase speeds of the two
wave modes are nearly identical. We thus expect a
transverse wave to change its amplitude along $\vect{N}$ and
$\vect{P}$ as it propagates, but not to show significant dispersion
between those two directions.

Finally, one should consider the meaning of the length coordinate of a
field line. A field line is not a physical entity, but merely a
convenient mathematical abstraction
\citep[e.g.,][]{2005LRSP....2....7L}.
At any given instant in time a length coordinate $s$ along the
field line can be unambiguously defined. We define $s=0$ to be at the
location where the field line crosses the $z=0$ plane and the magnetic
field vector $\vect{B}$ points upward (i.e, towards the corona), with
$s$ increasing in the direction of $\vect{B}$.  This coordinate system
does not remain fixed in time, because the field lines are moving with
the velocity field. The length of the infinitesimal line element
$\delta \mathbf{l}$ between two points $\mathbf{l}$ and
$\mathbf{l}+\delta \mathbf{l}$ lying on the same field line, with
$\delta \mathbf{l} =\delta s \mathbf{b}$ evolves as
\be 
\frac{d \ln \delta s}{dt} = \sum_{i,j} S_{ij} b_i b_j
\ee
where $b_i$ are the components of $\vect{b}$ and
$S_{ij}$ is the symmetric part of the velocity gradient tensor
$\partial u_i/ \partial x_j$
\citep[see for example][]{1992JFM...236..415D}.
In practice this means that one needs to be careful when interpreting
signal speed in $st$-diagrams of a field line as expansion and
contraction of the field line changes the apparent signal speed.

% *******************************************************************
\section{Analysis of fibril-threading fieldlines} \label{sec:analysis}
% *******************************************************************

% *******************************************************************
\subsection{Do fibrils trace field lines in 3D?} \label{sec:ffspace}
% *******************************************************************

% *******************************************************************
\begin{figure*} 
  \centering
  \includegraphics[width=14.2cm]{\figspath/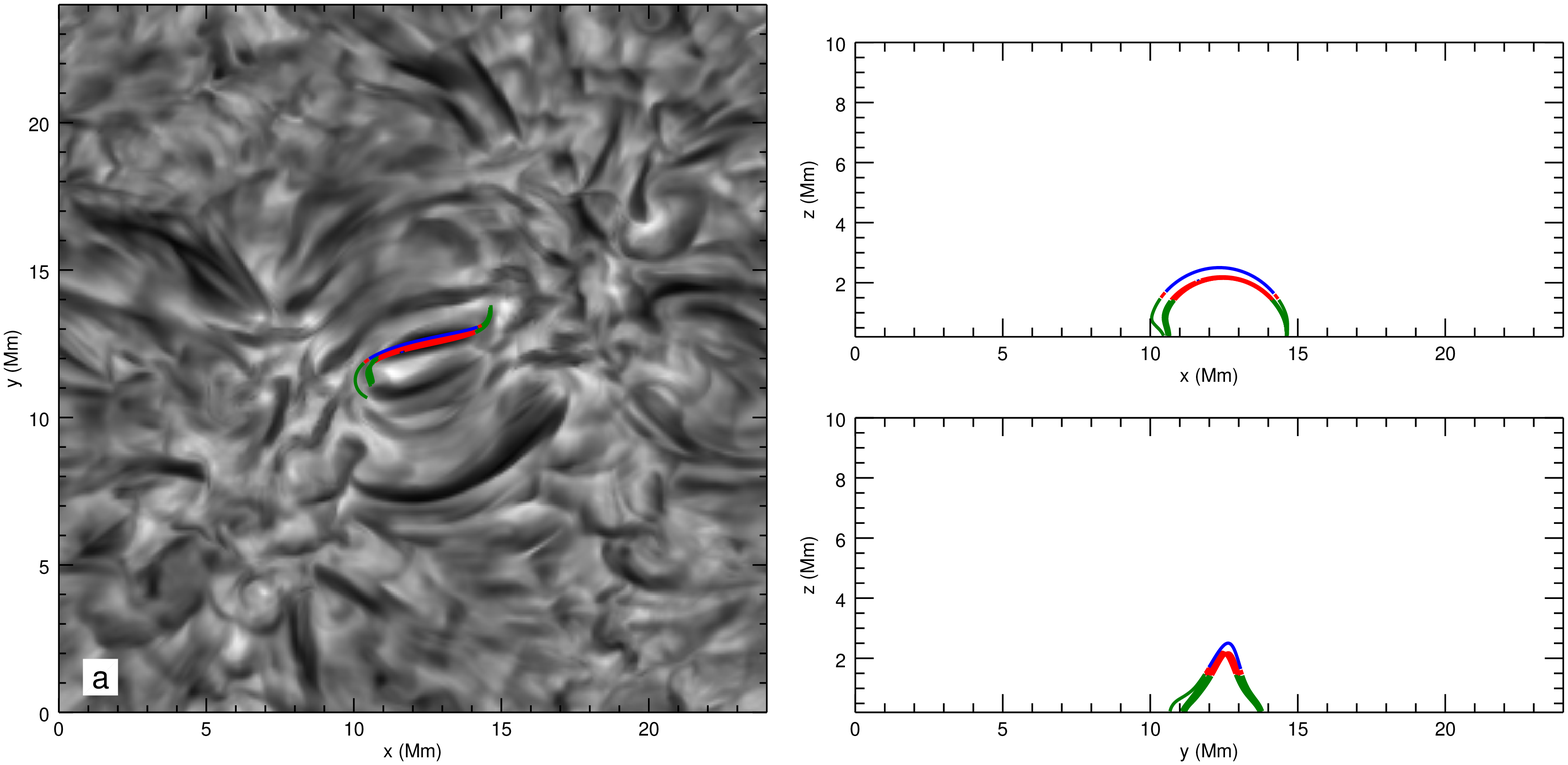}
  \includegraphics[width=14.2cm]{\figspath/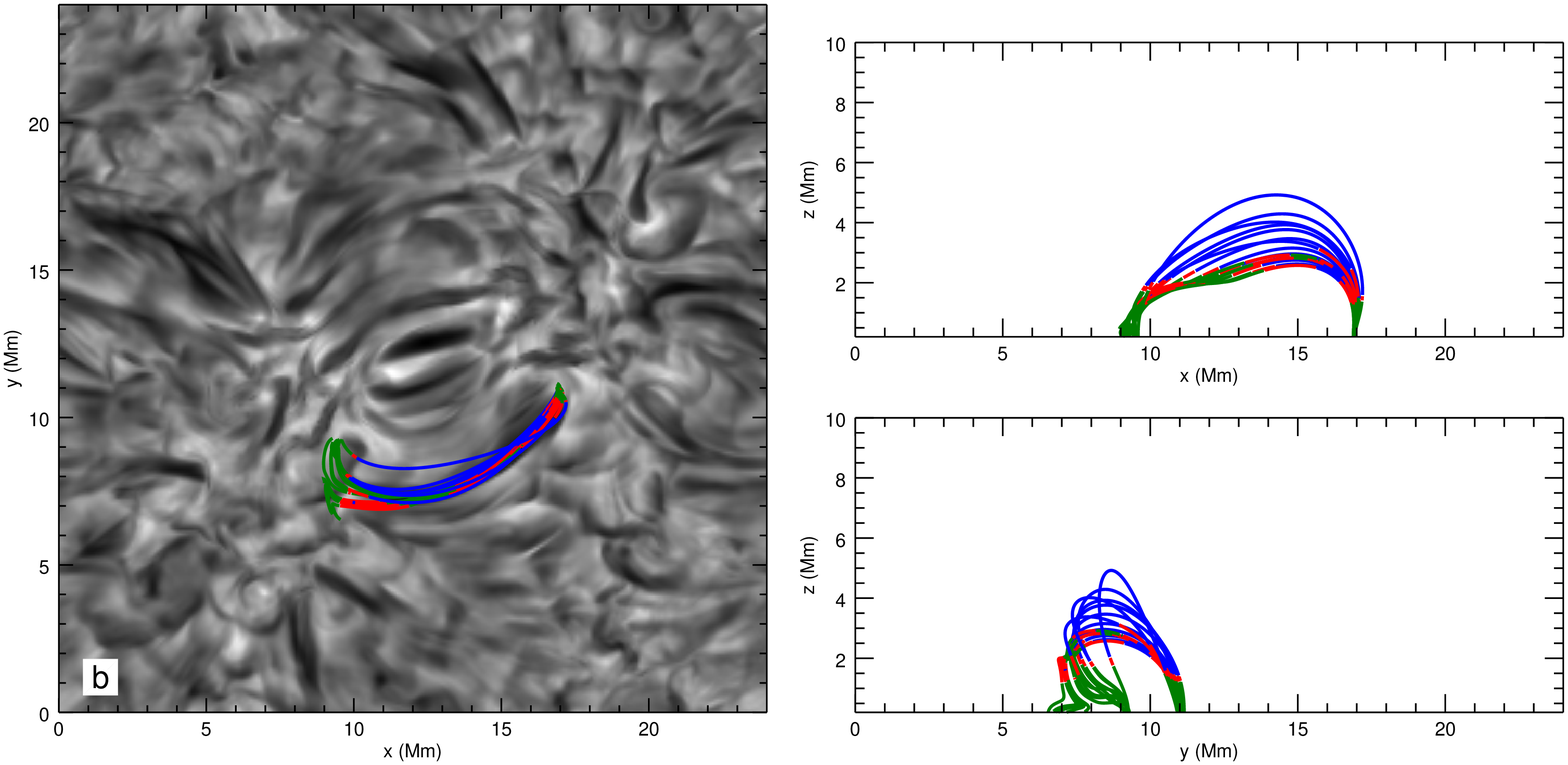}
  \includegraphics[width=14.2cm]{\figspath/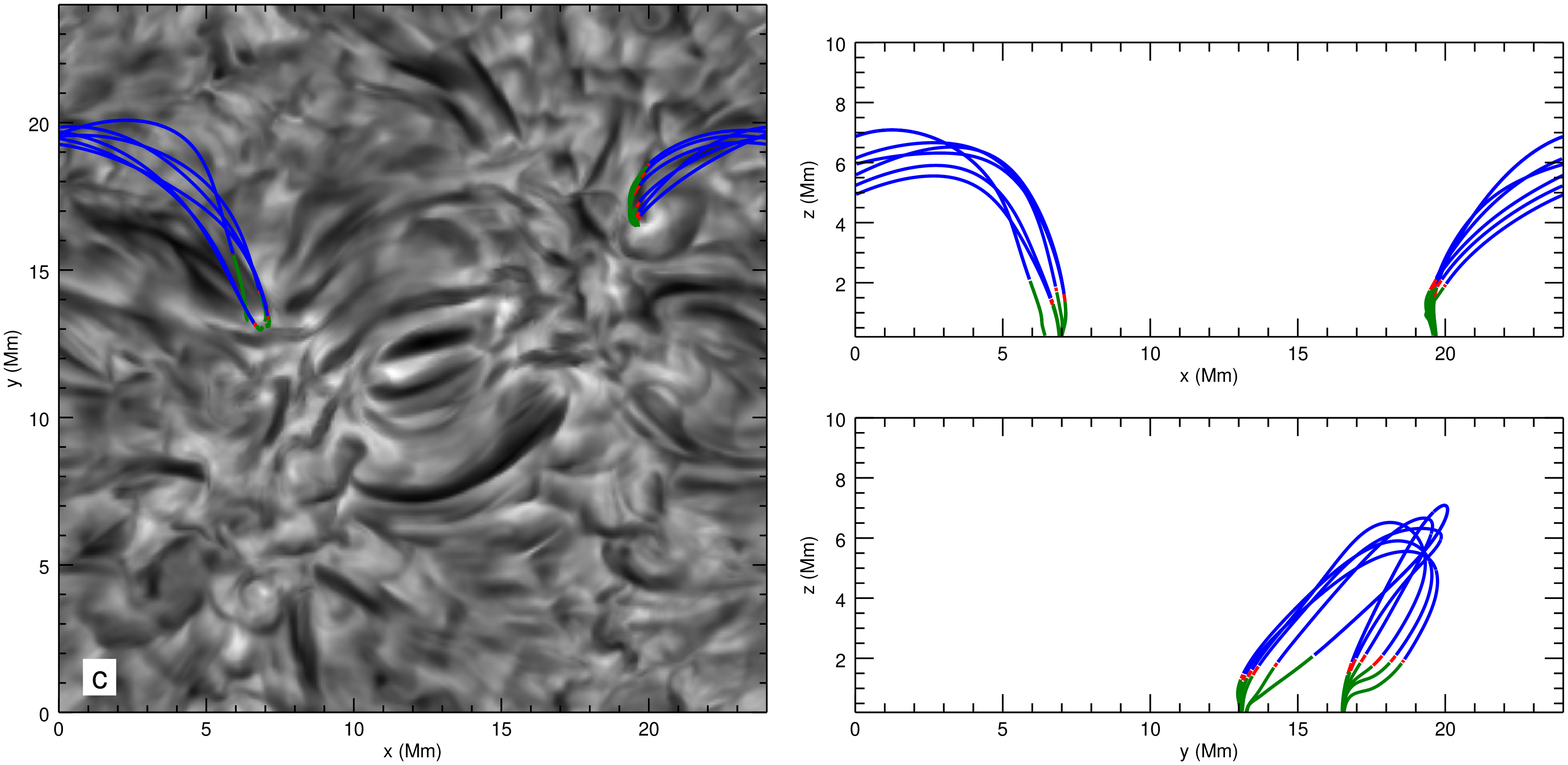}
   \caption{Visualization of flux bundles threading through three
     different fibrils at $t=600$~s. The labels a, b and c correspond to the labels
      in Figure~\ref{fig:flines}.
      The left panels show the top view of the field
     lines, superimposed on the \Halpha\ line-core image. The panels
     on the right side show the $xz$ and $yz$ side views. The field
     lines are colour coded: the part of the field lines more than 200
     km below optical depth unity in the line-core are green. The part
     more than 200 km above optical depth unity is blue, and the part
     within 200 km of optical depth unity is red. The red part of the
     field lines can thus be considered as visible in the
     \Halpha\ line core.  {\it A movie showing the time evolution of
       the fibrils and field lines can be found in the online
       material}.
  \label{fig:fibril_class}}
\end{figure*}
% *******************************************************************

In Figure~\ref{fig:flines} we show the seed points and associated
field lines that we selected for further analysis. The seed points
follow a number of fibrils, mostly very prominent ones, but also one
fibril that appears almost transparent at $t=600$ s but is more visible before
and after this moment. The $z$-coordinate of the seed point is the
height of optical depth unity at that $(x,y)$ location.

The middle panel shows the part of the field lines above the height of
optical depth unity. The field lines in the centre area 
remain low, while
the other field lines reach large heights. The bottom panel shows that
the field lines that point away from the centre of the domain typically
reach coronal temperatures, while the central ones have chromospheric
temperatures, with the latter defined as $T<17$~kK. The field lines
typically end in strong photospheric field concentrations, but there
are a few instances where one endpoint lies just next to a
concentration. 
These photospheric endpoints do not visually appear to be associated with a
fibril. 
The ending of fibril-threading field lines in photospheric
magnetic field concentrations is consistent with the observational
evidence that \Halpha\ fibrils are found as if emanating
from 
photospheric magnetic concentrations
\citep[e.g.,][]{2001SoPh..199...61C,2003A&A...402..361T,2007ApJ...660L.169R},
and further corroborates the idea that fibrils trace at
least the horizontal components of field lines that start in
photospheric magnetic field concentrations. Also note that the field
lines show a strong suggestion of twist and shear, indicating that the
simulated chromosphere is not in a force-free state. 

Figure~\ref{fig:fibril_class} displays for three different cases (a, b and c, also labeled in Figure~\ref{fig:flines}) how the fibrils and the field
lines intersect each other, and which part of the field lines can be
considered visible in \Halpha\ line-core images. The definition of
visible is necessarily somewhat imprecise, because the \Halpha\
contribution function is typically non-zero over a range of several
hundred kilometers, and can even have both a chromospheric and
photospheric component
\citep{2012ApJ...749..136L}.
We settled for defining a point $\mathbf{r}=(x,y,z)$ on a field line
as visible if it lies within 200 km from the height of optical depth
unity along the same column $z_{\tau=1}(x,y)$.

The top row of panels of Figure~\ref{fig:fibril_class} shows case a: a short
fibril in the centre of the computational domain. All field lines except one
form a very tight bundle, and are visible along the whole length of
the fibril. 
The exception is the highest-reaching field line, whose
seed point was on one endpoint of the fibril and can thus be
considered an intermediate case between fibril and non-fibril field
lines. For this fibril the assumption that fibrils trace the magnetic
field in all three dimensions is true. This finding is confirmed too
in the middle panel of Figure~\ref{fig:flines} where the field lines
can be seen to be almost parallel to the $\tau=1$ surface. The
temperature along the field lines is chromospheric.  Note that the
field lines show a characteristic sigmoid shape in the top view of Figure~\ref{fig:fibril_class}. 

In the middle row of Figure~\ref{fig:fibril_class} we show case b, an example
of a longer fibril that does not trace the magnetic field in three
dimensions. In the middle panel of Figure~\ref{fig:flines} the
field lines can be seen to thread through the optical depth unity
surface of the fibril at an angle. Different field lines are visible
at different locations along the length of the fibril. At one end of
the fibril at $(x,y)=(17,11)$ Mm the field lines all emanate from a
single magnetic element, whereas at the other end the field lines
connect to different magnetic elements in a patch of about
$1$ Mm$\times 3$ Mm. 
Also these field lines remain chromospheric and
do not protrude into the corona.

Finally, the bottom row of Figure~\ref{fig:fibril_class} shows case c: field
lines threading through a short fibril at $(x,y)=(20,17)$ Mm. This
case is very different from the previous two. The optical depth unity
surface in the fibril is nearly horizontal, and the field lines
intersect this surface at large angle (more than $45^{\circ}$). In the
$yz$-view at $y=17$ Mm the field lines are seen to form a fan-like
structure. The field line apexes reach transition region and coronal
temperatures as shown in the bottom panel of Figure~\ref{fig:flines}.

% *******************************************************************
\subsection{Time evolution of fibrils and field lines} \label{sec:fftime}
% *******************************************************************

We now turn our attention to the question whether fibrils trace the
same field line (or parts of field lines) over their lifetime. In the
supplemental material we provide movies of the time evolution of the
three fibrils and associated field lines shown in
Figure~\ref{fig:fibril_class}. In each case the fibrils and test
field lines intersect at $t=600$~s.  In all three cases the field
lines do not start out tracing a fibril at $t=0$~s, but instead seem
to ''drift'' into the fibril around $t=600$~s. After that time the field lines
drift out of the fibril, while the same apparent fibril remains
visible. Typically the field lines remain visible for about
100--200 s, which is of the same order as a typical observed wave
period.

% *******************************************************************
\subsection{Analysis of the dynamics of single field lines}
% *******************************************************************

% *******************************************************************
\begin{figure*} 
  \centering
  \includegraphics[width=15cm]{\figspath/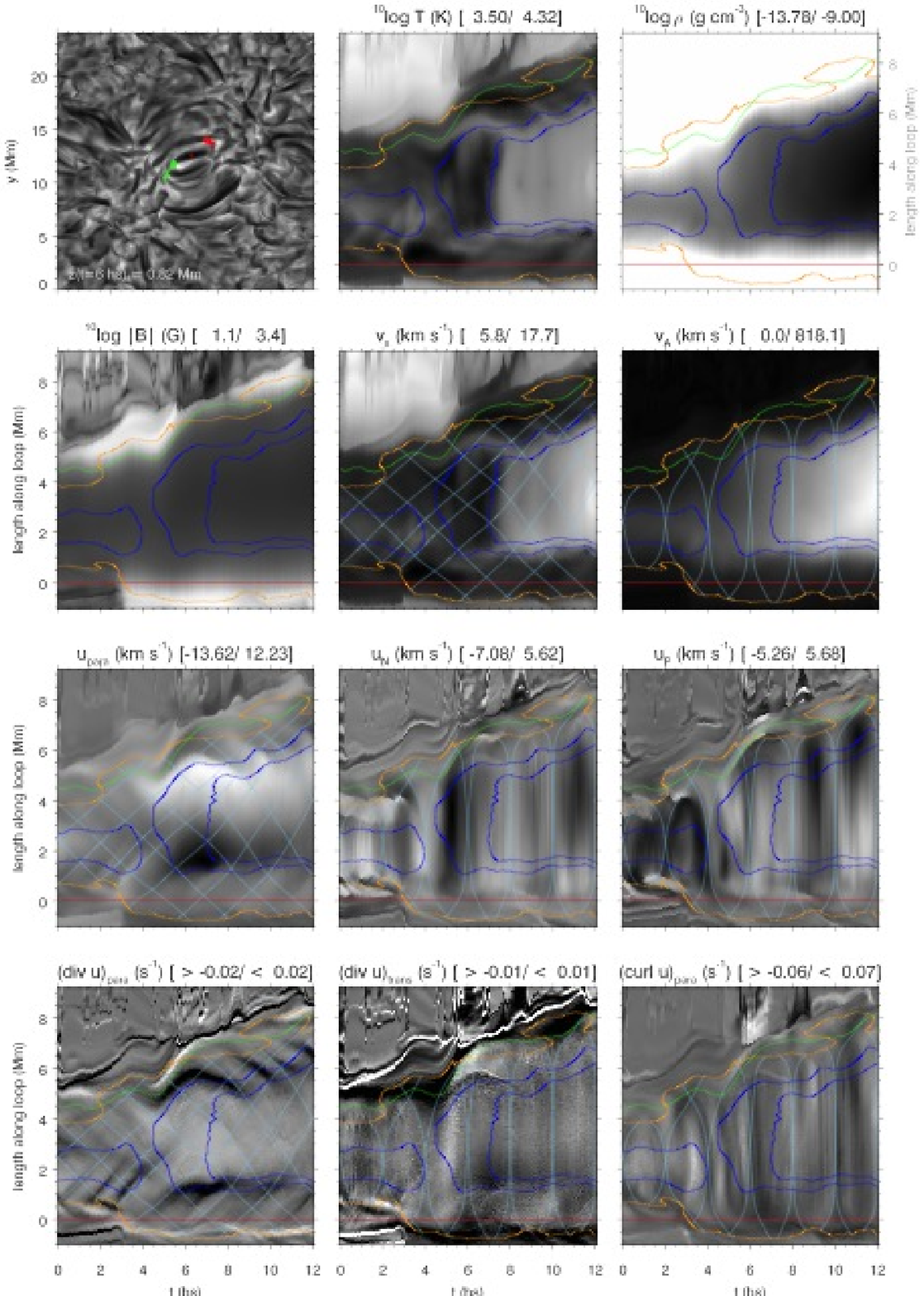}
   \caption{Evolution of fibril-threading field line A. The upper left
     panel shows a top view of the computational domain, with the
     location of the field line at $t=0$, $t=600$ and $t=1200$~s. The
     location and time evolution of the footpoints are indicated in
     red ($\vect{b}$ points up towards the corona) and green
     ($\vect{b}$ points down into the convection zone). The red cross
     marks the location of the original seed point from which the
     field line and its time evolution are computed. All other panels
     show $st$-diagrams: the time evolution (abscissa) of quantities
     as function of the loop length $s$ (ordinate). The displayed
     quantity and its range are given above each panel.  The zero
     point of $s$ is the intersection of the field line with the $z=0$
     plane where $\vect{b}$ points upwards. It is indicated with the
     red line. The green curve is the $s$-coordinate of the other
     footpoint. Areas of the field line within the dark blue curves
     are visible in the \Halpha\ line core. The orange curves
     indicates plasma $\beta=1$. The light blue curves are wave
     trajectories, i.e, a signal propagating with the wave speed along
     the loop would follow the indicated trajectory. The wave speed is
     the sound speed for the panels labeled $v_\mathrm{s}$, $v_\para$
     and $(\mathrm{div} \, u)_\para$ and the Alfv\'en speed for all
     other panels.
  \label{fig:fibril_050_st}}
\end{figure*}
% *******************************************************************

% *******************************************************************
\begin{figure*} 
  \centering
  \includegraphics[width=15cm]{\figspath/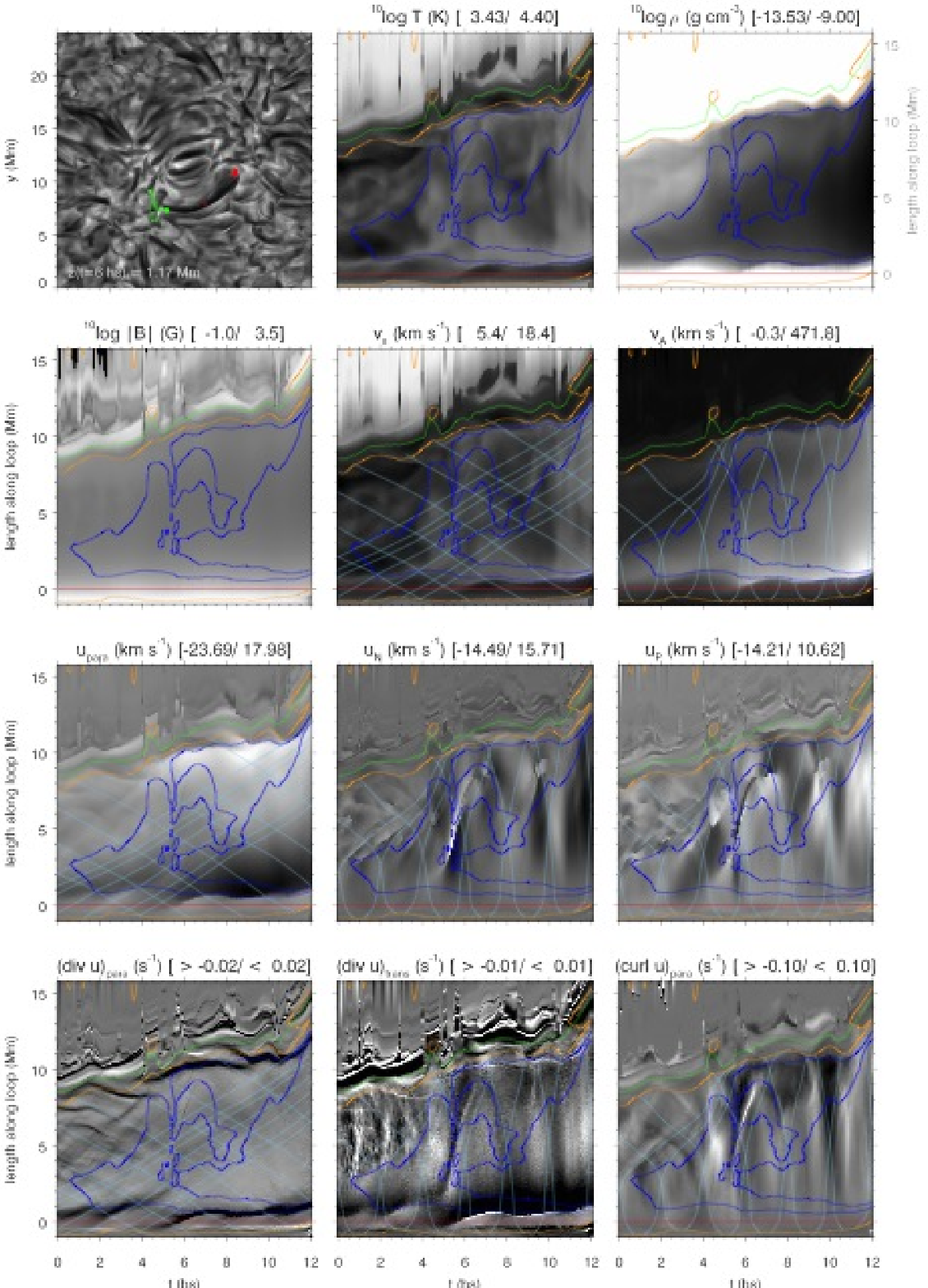}
   \caption{Same as Figure~\ref{fig:fibril_038_st} but for field line B.
  \label{fig:fibril_038_st}}
\end{figure*}
% *******************************************************************

% *******************************************************************
\begin{figure*} 
  \centering
  \includegraphics[width=15cm]{\figspath/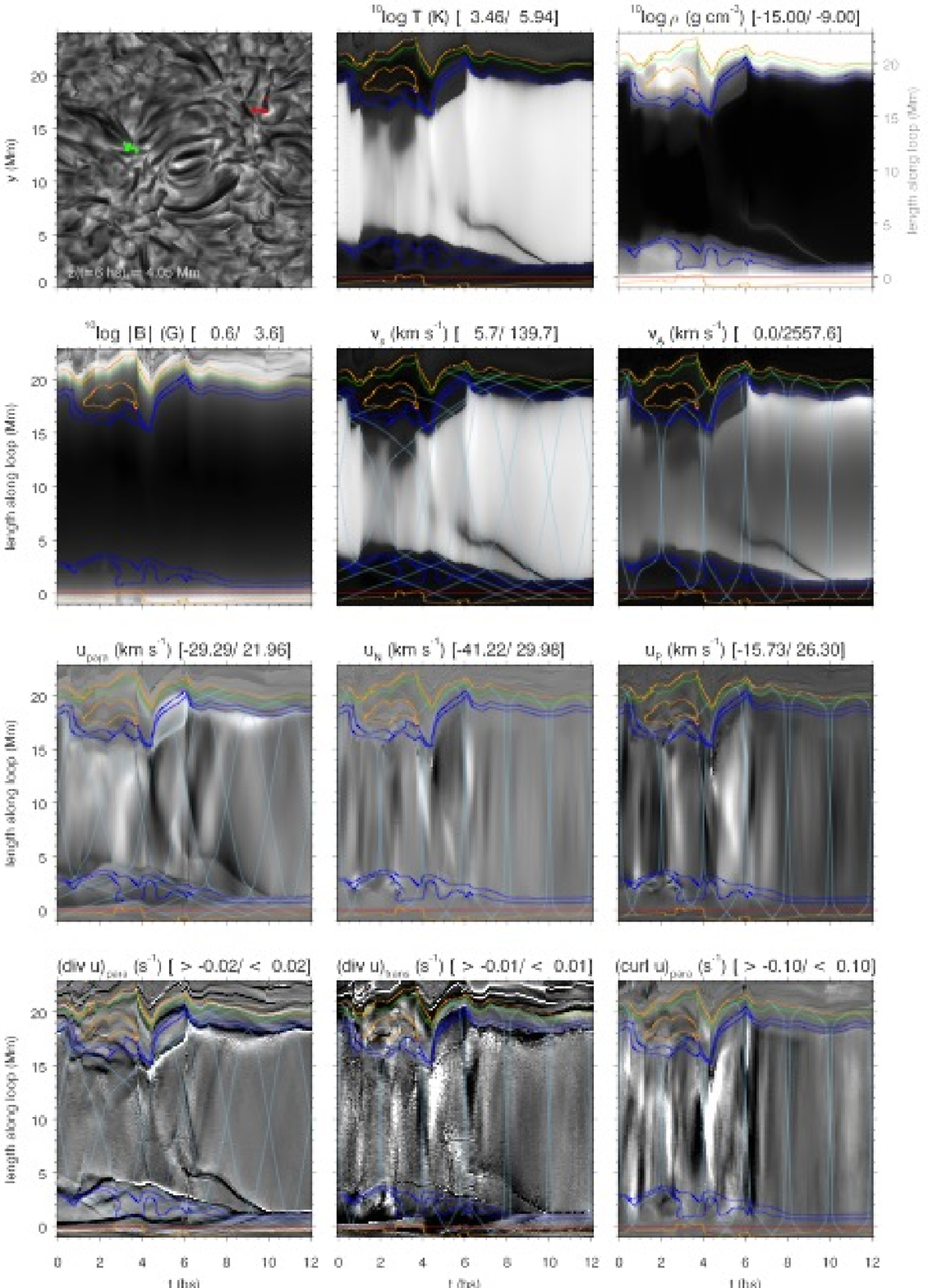}
   \caption{Same as Figure~\ref{fig:fibril_050_st} but for field line C.
  \label{fig:fibril_019_st}}
\end{figure*}
% *******************************************************************

Figures~\ref{fig:fibril_050_st}--\ref{fig:fibril_019_st} show the time
evolution of various quantities along the length $s$ of three
fibril-threading field lines, which we label A, B and C, one from each
of the three cases a,b, and c shown in Figure~\ref{fig:flines} and~\ref{fig:fibril_class}. The seed
points of these field lines were in the centre of the fibrils at
$t=600$~s.  In the online material we provide movies of the time
evolution of the 3D position of the three field lines.

The $st$-panels show the evolution of temperature
$T$ and density $\rho$ in the first row; magnetic field magnitude
$|\vect{B}|$, sound speed $v_\mathrm{s}$ and Alfv\'en speed
$v_\mathrm{A}$ in the second row; plasma flow speed parallel to the field
line $u_{||}=\vect{u} \cdot \vect{b}$, 
%LJ In the figures you call this u_para. Do that here too to avoid confusion?
and along the two
transverse directions $u_{\vect{P}}=\vect{u} \cdot \vect{P}$ and
$u_{\vect{N}}=\vect{u} \cdot \vect{N}$ in the third row. In the fourth
row we show the divergence of the plasma flow
parallel to the magnetic field 
\be
\nabla_{||} \cdot \vect{u} = \partial u_{||}/ \partial s,
\ee
 the divergence of the flow field
perpendicular to the field 
\be
\nabla_{\perp} \cdot \vect{u} = \nabla \cdot \vect{u} 
- \partial u_{||}/ \partial s,
\ee
and the vorticity parallel to the magnetic field 
\be
\omega_{||}=(\nabla \times \vect{u}) \cdot \vect{b}.
\ee

\subsection{Time evolution of purely chromospheric field lines} \label{sec:chromfl}

Figure~\ref{fig:fibril_050_st} shows the evolution of field line A
that forms a small loop in the centre of the field-of-view. The
\Halpha\ fibril at $t=600$ s traces the field lines in this loop very
well (See Section~\ref{sec:ffspace}). The $\beta=1$ curves indicate
that the magnetic pressure is dominant throughout the chromospheric
part of the field line. The blue curves, that indicate whether the
field line is visible show that the whole loop apex is visbile in
\Halpha\ for a duration $\simeq 200$ s around t=600 s. At earlier
times the field line is buried too deep in the atmosphere, and at
later times the loop has risen above the optical depth unity
surface. The field line has typical chromospheric temperatures between
3~kK and 15~kK. As it rises over time the density of the loop apex is
steadily decreasing. The magnetic field remains rather constant over
time. The sound speed just depends on the square root of the
temperature and varies between 5.8~and 17~\kms\ in the
chromosphere. In contrast, the Alfv\'en speed shows a much large
variation. At the apex of the field line, it increases over time from
50 kms to 800 \kms as the field line rises and its density
decreases. The light-blue wave-speed signal tracks illustrate
this. The footpoint-to-footpoint loop crossing time for sound waves is
more than 600 s, whereas the Alfv\'en crossing time is of the order of
100 s.

The $u_{||}$ and $\nabla_{||} \cdot \vect{u}$ panels reveal a pattern
of diagonal stripes from the $\beta=1$ curve towards the loop
apex. They are most clearly seen in the $\nabla_{||} \cdot \vect{u}$
panel as dark stripes, indicating compression of the plasma along the
direction of the loop. They are the signature of
photospherically-excited slow-mode waves that only propagate parallel
to the field lines in the low-$\beta$ regime.  Their slopes (and thus
propagation speeds) are comparable with the sound-speed, as can be
seen from comparing them with the sound-speed tracks. They are not
exactly equal because of loop expansion and contraction, variation in
space and time of the sound speed and the presence of background flows
which modify the wave speed in the observer's frame. Interestingly,
the slow-mode waves to not cross the loop from one end to the
other. Instead they typically fade before they reach the loop apex. A
preliminary analysis of these waves (not shown here) shows that the
radiative cooling time in the shock front is of the order of ~200
s. The waves are thus damped within a sound-speed crossing time.

The $u_{||}$ panel shows another interesting phenomenon. For
600~s $\lesssim t \lesssim 800$~s 
it shows a white patch around $s=5$ Mm and a black
patch around 
$s=2$ Mm. These are flows along the field line from the apex
towards both footpoints and thus represent the field line draining of
mass as it rises through the chromosphere.

Finally we look at the plasma motion orthogonal to the field line in
the panels $u_{\vect{P}}$ and $u_{\vect{N}}$. The field line is
approximately semi-circular, so $u_{\vect{P}}$ can roughly be thought
of as vertical oscillation, and $u_{\vect{N}}$ as horizontal
oscillation. Careful inspection of both panels shows a pattern of
perturbations propagating with the Alfv\'en speed. The perturbations
originate in the $\beta=1$ layer in both footpoints and travel all
along the field line to the other footpoint with little or no apparent
damping. There appears to be no wave reflection. The $\nabla_{\perp}
\cdot \vect{u}$ panel shows that these perturbations are only weakly
compressive. Only the perturbation around $t=5$ hs leaves a clearly
recognizable imprint of weak compression orthogonal to the field line.
The panel that shows $\omega_{||}$
exhibits a strong imprint of perturbations traveling with the Alfv\'en
speed and thus provide a hint of the presence of torsional
waves. However, true confirmation requires a more thorough
investigation of multiple field lines.

Field line B, displayed in Figure~\ref{fig:fibril_038_st}, shows by and
large the same features as Figure~\ref{fig:fibril_050_st}: the field
line remains at chromospheric temperatures, it carries slow-mode waves 
that are damped before they cross to the other footpoint and it
shows an interference pattern of transverse waves and possibly
torsional waves that propagate at Alfv\'en speed. The biggest
difference is the velocity amplitude of the transverse waves, it
reaches up to 15 \kms\ for this case, while field line A in
Figure~\ref{fig:fibril_050_st} showed a maximum amplitude of only
7 \kms.

\subsection{Time evolution of a coronal field line}

Field line C shown in Figure~\ref{fig:fibril_019_st} behaves
differently from the chromospheric field lines discussed in
Section~\ref{sec:chromfl}. The loop is $\sim20$ Mm long, twice the
length of the chromospheric field lines A and B. A large fraction of
the loop has coronal temperatures, and only a short part close to the
footpoints are chromospheric. The temperature and mass density 
panels show
the formation of a cold, dense condensation at $t=600$~s and
$s=6$~Mm. After its formation, it falls towards the chromosphere,
hitting it at $t=1000$~s. These properties are similar to what is
observed for coronal rain, even though the spatial scale in the
simulation is much smaller
\citep[see for example][and references
  therein]{2011ApJ...736..121A,2012SoPh..280..457A}.

Slow-mode waves propagating along this field line ($u_\para$ and
$\nabla_{||} \cdot \vect{u}$ panels) do travel the whole length of the
field line. This is caused by the quadratic density dependence of the
radiative cooling. 
The wave crosses the chromosphere too fast
to be completely damped. 
Once the wave front reaches the corona, the
radiative cooling timescale goes down dramatically and becomes longer
than the wave crossing time. \edt{Both loop footpoints are sources of slow-mode waves, and their lack of damping in the corona leads to an interference pattern in $u_\para$ and $\nabla_{||} \cdot \vect{u}$, but the latter is not as clear in Figure~\ref{fig:fibril_019_st} because of their small compression compared to the slow-modes in the chromosphere.}

The $u_{\vect{P}}$ and $u_{\vect{N}}$ again show transverse
oscillations running at Alfv\'en speed along the field line 
\edt{in both directions}.
The
vorticity panel ($\omega_{||}$) shows a
hint of torsonial waves. 

Note that the Alfv\'en crossing time for this coronal field line is
$\sim40$~s assuming a length of 20 Mm and a speed of 500 \kms,
so that we barely resolve the propagation at the 10 s cadence of our
time series.

\subsection{Velocity power spectrum}

% *******************************************************************
\begin{figure*} 
 \includegraphics[width=8.8cm]{\figspath/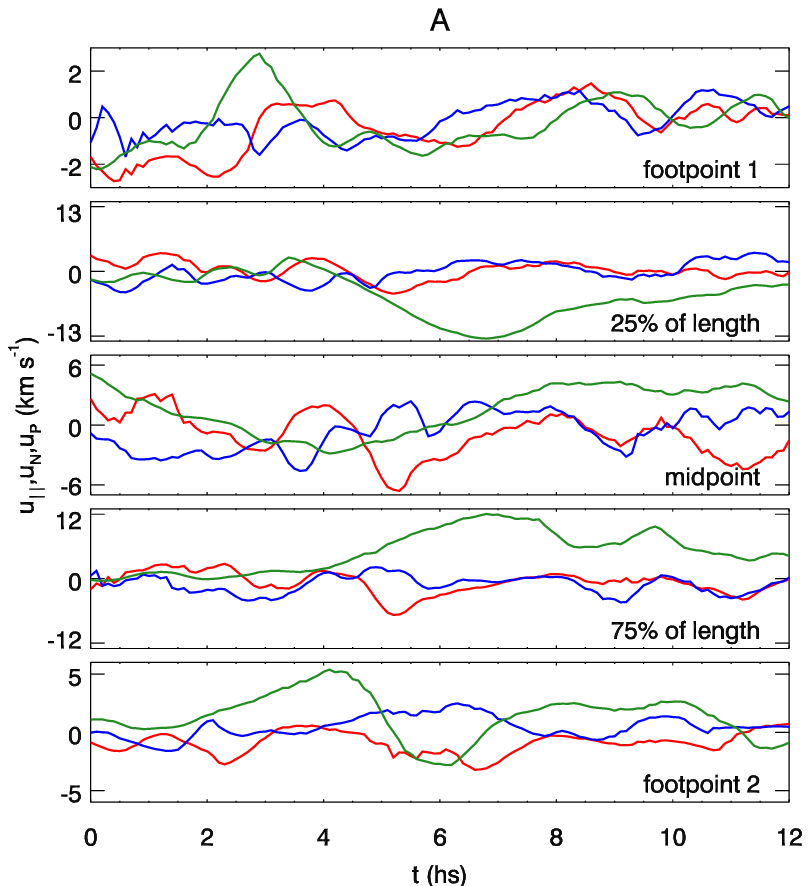}
 \includegraphics[width=8.8cm]{\figspath/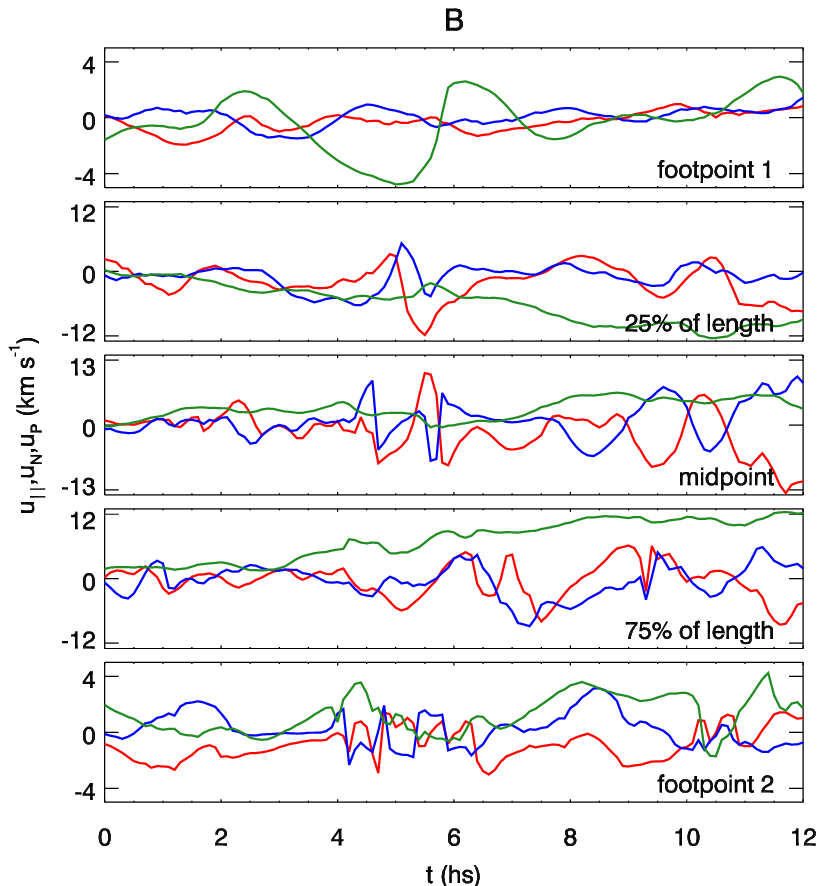}
 \includegraphics[width=8.8cm]{\figspath/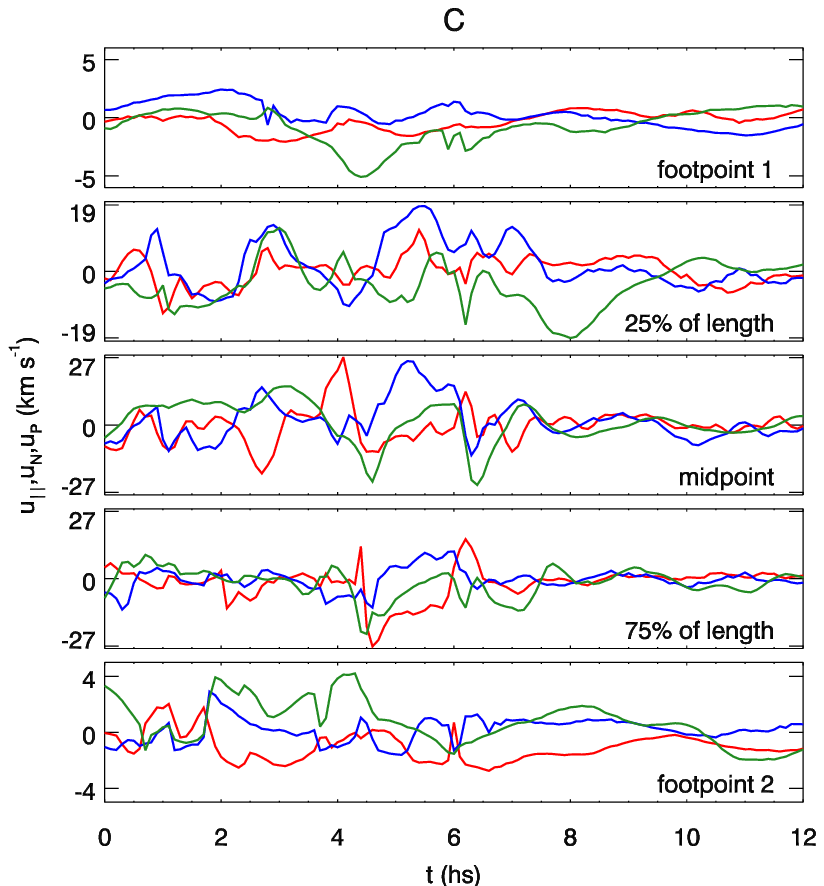}
  \includegraphics[width=8.8cm]{\figspath/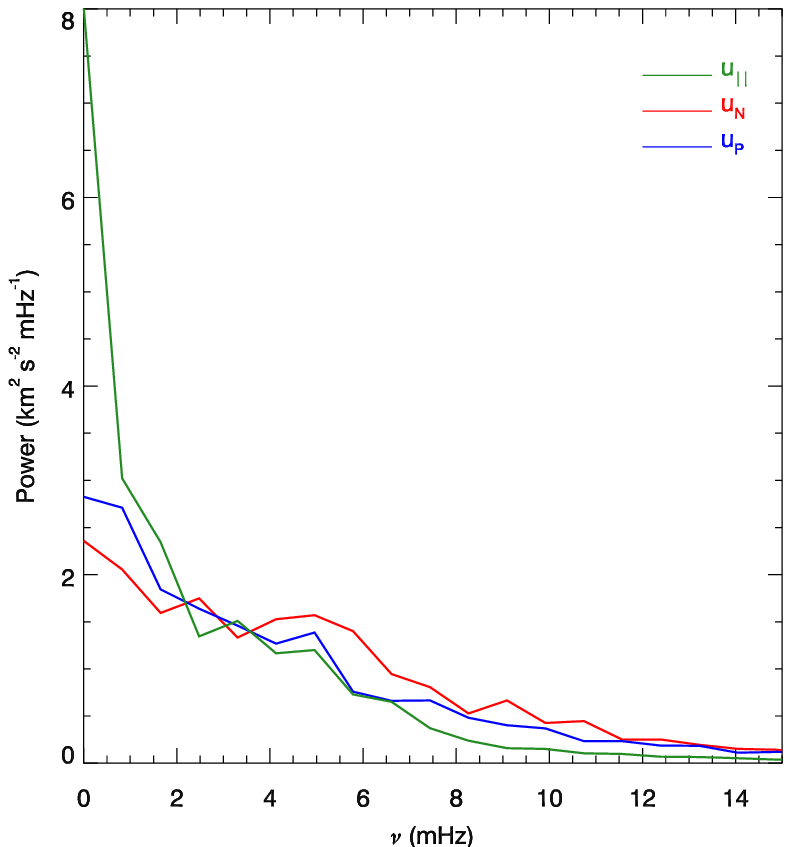}
   \caption{Top row and bottom left: Velocities as function of time
     parallel ($u_{||}$, green) and perpendicular to the magnetic
     field ($u_N$ in red and $u_P$ in blue) along the three field
     lines A,B,C. Subpanels show the velocities at different locations
     along the field line, where 100\% means the length from one
     footpoint to the other. Lower right panel: power
     spectrum of the velocity components at the loop apexes averaged
     over all 80 field lines. Only positive frequencies are
     shown.
    \label{fig:power_spect}}
\end{figure*}
% *******************************************************************

For completeness we further quantify the flow velocities along the
field lines A, B and C in Figure~\ref{fig:power_spect}. The velocities
increase from the footpoints towards the field line midpoints. The
parallel and perpendicular velocities have similar magnitude at
similar locations along the field lines. The parallel velocity
$u_{||}$ varies on longer time scales than the two perpendicular
components. This slow variation is associated with field lines loading
and draining of mass.

In the lower right panel of Figure~\ref{fig:power_spect} we show the
power spectrum of the three velocity components at the midpoint of the
field lines (i.e., halfway between the footpoints). These power
spectra were obtained by computing a power spectrum of each velocity
component at 11 points equally spaced between 40\% and 60\% of the
length of each field line. For each velocity component these spectra
were then averaged over the 11 points and over all field lines. The
averaging over 11 points was done to beat down the noise.

The high power of $u_{||}$ at the lowest resolvable frequencies again
represents loop mass loading and draining. There are weak maxima of
power at 5~mHz in all three velocity components. Given the limited
sample size and duration of the time series, we do not know whether
these peaks are significant. If they are, they might correspond to the
familiar 3-minute oscillations. The typical power of 1-2 km$^2$
s$^{-2}$ mHz$^{-1}$ are the same order of magnitude as those found
based on fibril oscillations in Figure~4 of
\citet{2014ApJ...784...29M}.

\section{Discussion and conclusions} \label{sec:discussion}

In this paper we investigated the relation between fibrils, dark elongated features seen in images of the Sun taken in the core of the chromospheric \Halpha\ line, and magnetic field lines that thread through the chromosphere.
We compared synthetic \Halpha\ imagery
computed from an RMHD simulation of an enhanced network area on the
Sun with observations. The simulation behaves similar in terms of
intensity contrast, Doppler shifts and time evolution. A difference is
that the fibrils in the synthetic imagery appear as a cross between
active region and quiet Sun fibrils. Comparison of the properties of
the oscillatory behaviour of fibrils in the synthetic line core
imagery with those observed by
\citet{2012ApJ...750...51K,2013ApJ...779...82K}
and
\citet{2014ApJ...784...29M}
showed that the simulated fibrils have periods, amplitudes, and phase
speeds consistent with the observations.

We then proceeded with an extensive analysis of the relation between
fibrils and fibril-threading field lines in our simulation. This
relation is one of complex interplay between field line dynamics on
the one hand, and the formation of the \Halpha\ line on the other
hand.

The evolution of the field lines themselves is extremely
intricate. 
The slow motion of their footpoints by the evolving granulation
pattern leads to a slow migration of the field line paths through the
3D computational volume on time scales of several hundred seconds. In
our examples, the field lines tend to evolve from lower
to
higher-reaching loops. The field lines can in general thus not be
thought of as a static background along which waves propagate.

Footpoint buffeting by radial box oscillations (corresponding to
solar $p$-modes) and granules cause excitation of different types
of waves. 

We identified the presence of compressive longitudinal slow-mode
waves, that steepen to shocks at chromospheric heights and propagate
with speeds close to the sound speed.  The plasma $\beta$ is smaller
than unity along the field lines, so that these waves propagate
parallel to the field lines. 
% Our analysis shows that they can fill fibril-threading field lines with mass. 
If these field lines are close
to horizontal, then the large angle of the field lines with the
vertical combined with the frozen in condition might prevent the loop
from draining quickly. For those field lines that do not reach coronal
temperatures, the slow-modes are typically damped before the wave
front reaches the other footpoint. This damping is reduced once the
shock front reaches to the corona, and such waves travel through the
corona to the other footpoint.

The second wave type that we identified is a nearly-incompressible
transverse wave that propagates with the Alfv\'en speed. These waves
are ubiquitously present in all investigated field lines along both
transverse Frenet-Serret directions ($\vect{N}$ and $\vect{P}$). They
have maximum velocity amplitudes of 5--20 \kms. The field lines
typically carry waves travelling in both directions along the field.
We do not find evidence of standing waves or reflection. We also find
signal propagating with Alfv\'en speed in the vorticity parallel to
the magnetic field $\omega_{||}$.  Without further analysis we cannot
confirm whether the vorticity signal indicates a torsional wave-like
phenomenon, such as reported in
\citet{2012ApJ...752L..12D,2014Sci...346D.315D}
or whether it is caused by a form of motion comparable to helical kink
waves in idealised linear wave analysis
\citep{2008ApJ...683L..91Z}.
In the latter case they merely represent another way to view the
transverse waves along $\vect{N}$ and $\vect{P}$.

Our field line based analysis approach only brings forward waves that
travel parallel to the magnetic field. The third classical MHD wave,
the fast mode, can propagate at an angle to the magnetic field  \edt{and can therefore not be easily identified in our field-line based approach}. While
these waves are present in the simulation, we did not try to identify
them.

From an \Halpha\ line formation point of view, the fibrils are dark
elongated features. They appear dark because they reach optical depth
unity at larger height than their surroundings. The \Halpha\ opacity
is relatively insensitive to temperature, and depends mainly on mass
density. The location of optical depth unity in the line core is
therefore roughly equal to the height were the integrated mass density
(column mass) along the line of sight reaches a certain value 
($3\times 10^{-5}$~g~cm$^{-3}$ in the analysis by
\citet{2012ApJ...749..136L}).
Whether (a part of) a given magnetic field line is visible as part of
a fibril therefore depends on the mass density along the field line,
and the column mass density above it. The column mass density at which
optical depth unity is located is just set by the characteristics of the
radiative transfer in the \Halpha\ line, and has no special
significance in terms of properties of the atmosphere.

The combination of the movement of field lines through the atmosphere
and the filling and draining of mass along field lines thus determines
whether a part of a given field line is visible in \Halpha\ line core
images. From the analysis of the time evolution of the fibrils and
field lines in the movies associated with
Figure~\ref{fig:fibril_class} we find that a fibril in our simulation
tracks the same set of field lines for about 200~s. The fibrils
themselves keep their identity typically for a longer time (500~s to
1000~s). In this respect our simulation behaves more like the active
region fibrils analysed in Figure~2 of
\citet{2014ApJ...784...29M},
where fibrils retain their identity for about 500~s, whereas the quiet
Sun fibrils are identifiable over shorter periods
(200~s\,--\,300~s). A big open question is whether true solar field
lines migrate through the atmosphere as fast as in our simulation.

Returning to the assumptions behind using fibril oscillations in
coronal seismology in Section~\ref{sec:introduction}, we find the
following: The simulation shows instances where fibrils outline single
magnetic field lines (top row of Figure~\ref{fig:flines}, but also
examples where the fibril 
roughly follows a field line bundle,
but different field lines are visible at different locations along the
fibril (middle row of Figure~\ref{fig:flines}) and examples where the
fibril intersects field lines at a large angle (bottom row of
Figure~\ref{fig:flines}). In the latter case the fibril does not
follow single field lines at all. Whether a given fibril traces single
field lines cannot be determined from the \Halpha\ observations alone.
The simulation also shows that field lines evolve in time though the
computational domain, and do not intersect a fibril for longer than $~
200$~s. This is comparable to typical transverse wave periods.
 
We conclude that our simulation only partially supports the
assumptions typically made in chromospheric seismology. There will be
instances where a fibril indeed traces out a single field line over
more than one transverse-wave period, but this cannot be expected to
be generally true. We therefore urge caution when interpreting results
obtained from seismology, as they might contain systematic biases
caused by the partial mapping of fibril oscillations to actual plasma
and field line motion.

The work presented in the manuscript also leads to a number of
intriguing new questions: 
What are the precise excitation mechanisms
of the transverse waves along fibril threading field lines? How can
fibrils retain their identity for a time that is longer than the time
that they intersect the same field lines? How much of the transverse
and longitudinal wave energy is deposited in the chromosphere? What is
the dissipation mechanism of the waves? How exactly is mass loaded 
into fibril-threading field lines? Which forces act to prevent
the mass-loaded fibrils to sink back deeper into the atmosphere? Why
are fibrils in active regions narrower than in the quiet Sun? We
intend to address these questions in future publications.

%_______________________________________________________________________
\begin{acknowledgements}
 The research leading to these results has received funding from the European Research Council 
 under the European Union's Seventh Framework Programme (FP7/2007-2013) / ERC Grant 
 agreement nr. 291058. This research was supported
 by the Research Council of Norway through
 the grant ``Solar Atmospheric Modelling'' and through grants of
 computing time from the Programme for Supercomputing. It was also
 supported by the Swedish Knut and Alice Wallenberg foundation. The
 Swedish 1-m Solar Telescope is operated by the Institute for Solar
 Physics of Stockholm University in the Spanish
 Observatorio del Roque de los Muchachos of the Instituto de
 Astrof\'{\i}sica de Canarias. The authors recognize support from the International Space Science Institute in Bern.
\end{acknowledgements}
%_______________________________________________________________________

\end{document}